\documentclass[12pt]{article}
\usepackage{a4wide}
\usepackage{cite}
\usepackage{latexsym}
\usepackage{epsfig}
\def\bq{\begin{eqnarray}}
\def\eq{\end{eqnarray}}
\def\l{\langle}
\def\r{\rangle} 
\def\eps{\varepsilon}

\def\g{\gamma}

\def\ifm{\ifmmode}
\def\msb{\ifm \overline{\rm MS}\, \else $\overline{\rm MS}\, $\fi}

\newcommand{\smin}{s_{\rm min}}

\newcommand{\rowspace}{\mbox{\rule[0pt]{0pt}{14pt}}}

\begin{document}

\thispagestyle{empty}

\begin{flushright}
RMU-SEMS-011002\\
NIKHEF/2002-005\\
ITP-2002/29\\
FERMILAB-Pub-02/134-T\\
UPRF-2002-10
\end{flushright}

\vspace{1.5cm}

\begin{center}
 {\Large\bf The fully differential single-top-quark\\[0.3cm]
cross section in next-to-leading order QCD}\\[1cm]
  {\sc  B.~W. Harris$^{\,a}$, E. Laenen$^{\,b,\,c}$, L. Phaf$^{\,b}$,
  Z. Sullivan$^{\,d}$, S. Weinzierl$^{\,e}$}\\
  \vspace{1cm}
  $^a${\it Robert Morris University, Coraopolis, PA 15108, USA} \\[0.2cm]
  $^b${\it NIKHEF, P.O. Box 41882, NL-1009 DB, Amsterdam,
    The Netherlands} \\[0.2cm]
  $^c${\it Institute for Theoretical Physics, Utrecht University,
    Utrecht, The Netherlands} \\[0.2cm]
  $^d${\it Theoretical Physics Department, Fermi National Accelerator
   Laboratory, \\ Batavia, IL 60510-0500, USA} \\[0.2cm]
  $^e${\it Dipartimento di Fisica, Universit\`{a} di Parma, \\
       INFN Gruppo Collegato di Parma, 43100 Parma, Italy} \\[0.2cm]
\end{center}

\vspace{.5cm}

\centerline{(July 3, 2002)}

\vspace{1.5cm}

\begin{abstract}\noindent
We present a new next-to-leading order calculation for fully
differential single-top-quark final states.  The calculation is
performed using phase space slicing and dipole subtraction methods.
The results of the methods are found to be in agreement.  The dipole
subtraction method calculation retains the full spin dependence of the
final state particles.  We show a few numerical results to illustrate
the utility and consistency of the resulting computer implementations.
\end{abstract}

\vspace*{\fill}

\section{Introduction}
\label{sec:introduction}

Single-top-quark production provides an excellent opportunity to study
the charged-current weak-interaction of the top quark
\cite{Willenbrock:cr,Yuan:1989tc,Cortese:fw,Ellis:1992yw,Carlson:1993dt,Stelzer:1995mi,Heinson:1996zm,Stelzer:1998ni}.
Measurement of the production cross section
of single top quarks is planned at 
both the Fermilab Tevatron and the CERN Large Hadron Collider (LHC)
\cite{Beneke:2000hk}.  The D0 \cite{D0} and CDF \cite{CDF}
collaborations have already set limits on both the $s$-channel and $t$-channel
cross sections using data collected during run I of the Fermilab
Tevatron, and have developed strategies for discovery at the current
run II.

Within the standard model, the single-top-quark cross section allows a direct
measurement of the Cabibbo-Kobayashi-Maskawa (CKM) matrix element $V_{tb}$.
Further, the short lifetime of the top quark presents an opportunity to
observe the polarization of the top quark at production, and hence
directly probe the $V-A$ nature of the weak interaction
\cite{Mahlon:1996pn,Mahlon:1999gz}.

Further, the mass of the top quark is comparable to the electroweak symmetry
breaking scale, which has lead to speculation regarding new physics
involving the top quark for some time \cite{Peccei}.
There are many proposals to use
these measurements to study non-standard couplings
\cite{Carlson:1994bg,Kane:1991bg,Rizzo:1995uv,Tait:1996dv,Datta:1996gg,Whisnant:1997qu,Boos:1999dd,Tait:2000sh,Espriu:2001vj},
strong dynamics
\cite{Simmons:1996ws,Lu:1997aa,Baringer:1997wu,Yue:ff}, flavor changing
neutral currents \cite{Tait:1997fe,Han:1998tp}, CP violation
\cite{Atwood:1996pd,Bar-Shalom:1997si,Christova:2001zm}, supersymmetry
\cite{Li:1996bh,Li:1996ir,Li:1997qf}, $R$-parity-violating
supersymmetry
\cite{Datta:1997us,Oakes:1997zg,Hikasa:1999wy,Chiappetta:1999cd}, and
Kaluza-Klein modes \cite{Datta:2000gm}.

The characteristics of single-top-quark production form a new class of
benchmarks for testing Quantum Chromodynamics (QCD).  Experimental
comparisons to the calculated kinematic distributions can provide a
handle on inputs to the calculation.  For instance, this process
involves directly the bottom quark parton distribution function.
Currently this function is constrained only indirectly though global
fits \cite{Lai:1997mg,Lai:1999wy,Martin:1999ww} to data.  The
bottom-quark density is then calculated from the light parton
densities \cite{Collins:1986mp,Aivazis:1993pi,Buza:1996wv,Smith:1997pi,Chuvakin:1999nx,Thorne:1997uu,Thorne:1997ga} in the context of QCD.
Including a process that depends directly on the bottom density into
the global analysis would be of great value to LHC observables, as
their typical scale will be much larger than the bottom mass where the
evolution begins.

Finally, single-top-quark production is a background to all signals
with $W+\mathrm{jet}$ or $W+b$ as backgrounds.  This background is
significant in a number of Higgs search channels (for a review see
\cite{Carena:2000yx}) and other new physics, such as $R$-parity
conserving \cite{Abel:2000vs} and violating 
\cite{Berger:1999zt,Allanach:1999bf} supersymmetry searches.

In this paper we present and discuss new calculations of
single-top-quark production at next-to-leading order (NLO) in QCD.
Earlier calculations of the NLO single-top-quark-production cross
sections exist in the literature.  The first calculation
\cite{Bordes:1995ki} was for the double differential cross section
and used small masses for the gluons and quarks 
to regularize infrared and collinear divergences.   Mass factorization 
was performed in the Deep Inelastic Scattering (DIS) scheme.
Subsequent NLO calculations for the $s$-channel
\cite{Smith:1996ij,Mrenna:1997wp} and $t$-channel
\cite{Stelzer:1997ns,Tait:1997fe} modes used dimensional
regularization and expressed the semi-inclusive cross section in terms
of the Modified Minimal Subtraction ($\overline{\rm MS}$)
factorization scheme.  The value of our work is that the results are
fully differential (meaning experimental cuts and jet finders can be
applied), the results contain spin information, and the results use
standard methods and schemes.  The calculational methods used are an
instructive step toward the computation of NLO corrections to the
$t\bar b j$ production channel that should also be considered when
studying single-top-quark production.

The NLO cross section receives contributions from virtual
corrections and real emission diagrams.  Taken separately, both parts
are divergent and therefore cannot be evaluated in a straightforward
way numerically on a computer.  Only the sum of the virtual
corrections and the real emission contributions is finite after mass 
factorization.  Writing a general-purpose NLO Monte-Carlo based
program therefore requires the analytic cancellation of singularities
before any numerical integration can be performed.

The two main general methods to handle the cancellation of
singularities without loss of information are the phase space slicing
\cite{Fabricius:1981sx,Kramer:1986mc,bergmann,boo1,Giele:1992vf,Giele:1993dj,Keller:1998tf,Harris:2001sx}
and the subtraction
\cite{Ellis:1980wv,kn,Kunszt:1992tn,Mangano:jk,Frixione:1996ms,Catani:1997vz,Phaf:2001gc,Catani:2002hc}
methods.  In this paper we implement the phase space slicing method of
one \cite{Keller:1998tf} and two \cite{Harris:2001sx} cutoffs, and the
subtraction-based dipole formalism of Ref.~\cite{Phaf:2001gc}.  We find
that the results of all methods agree.  The dipole calculation uses
helicity amplitudes and therefore contains the complete spin
correlations of the participating partons.

We organize this paper as follows.  In the next section we present an
overview of the amplitudes entering the calculation.  Section
\ref{sec:slicing} discusses the framework of the phase space slicing
method and presents the attendant analytic results.  Section
\ref{sec:dipole-subtraction} discusses the cancellation of divergences
within the context of the massive dipole subtraction method, and gives
the results for all relevant amplitudes.  This Section also contains a
detailed discussion of scheme independence for different ways of
handling $\gamma_5$.  The analytic results of Secs.~\ref{sec:slicing}
and \ref{sec:dipole-subtraction} are presented using different, but
self-consistent, notations appropriate to their methods of handling of
infrared divergences.  Brief numerical results are presented in
Sec.~\ref{sec:numerical}, along with a comparison of the methods.  Our
conclusions are followed by an Appendix containing the relevant
one-loop scalar integrals.

\section{Overview}
\label{sec:overview}

\begin{figure}[hbtp]
\begin{center}
\centerline{\hbox{\epsfig{figure=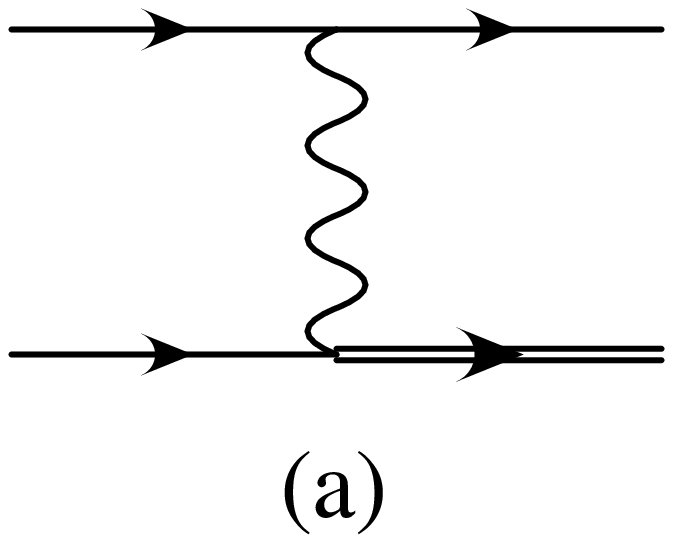,width=1.0in}\hspace*{.2in}
                  \epsfig{figure=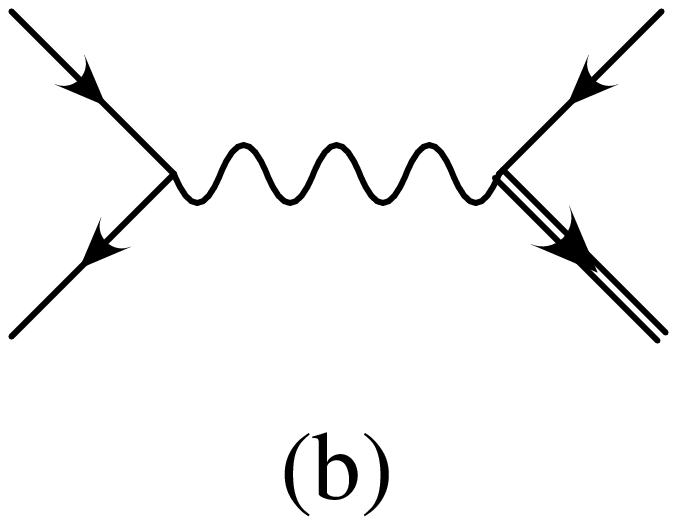,width=1.0in}}}
\caption{\small The leading order Feynman diagrams for reactions given in (a) 
Eq.\ \ref{eq:3} and (b) Eq.\ \ref{eq:6}.  The top quark line is doubled.}
\label{fig:1}
\end{center}
\end{figure}

The lowest-order Feynman diagrams are shown in Fig.~\ref{fig:1}.
They are commonly distinguished by the sign of the $W$ boson momentum squared.
The $t$-channel flavor excitation process
\begin{eqnarray}
\label{eq:3}
u + b \rightarrow t + d\,,
\end{eqnarray}
occurs via the exchange of a virtual space-like $W$-boson, and the
Drell-Yan-like $s$-channel process
\begin{eqnarray}
\label{eq:6}
u + \bar{d} \rightarrow t + \bar{b}\,,
\end{eqnarray}
occurs via a virtual time-like $W$-boson.  In reaction (\ref{eq:3}) it
is understood that we may replace the $(u,d)$-quark pair by
$(\bar{d},\bar{u})$, $(c,s)$ and $(\bar{s},\bar{c})$. In reaction
(\ref{eq:6}) we may replace the $(u,\bar{d})$-pair by $(c,\bar{s})$.
In addition, CKM suppressed combinations are included at all vertices.

For each channel the fully differential spin-averaged Born cross
section can be written as
\begin{equation}
  \label{eq:47}
  d\sigma_p^{(0)} =  \frac{1}{2s} \overline{\sum}
\left|{\cal M}_{p,\rm Born}\right|^2d \mathrm{PS}_2 \,,
\end{equation}
where $s$ is the partonic center-of-momentum energy squared, and we
use $p = s,t$ to denote the channel.  The two body phase space is
given by
\begin{equation}
  \label{eq:48}
  d\mathrm{PS}_2 = \frac{1}{(2\pi)^2}
    \frac{d^3 p_1}{2 E_1} \frac{d^3 p_2}{2 E_2}
   \delta^{(4)}(q-p_1-p_2)\, .
\end{equation}
The $t$-channel Born matrix element squared summed (averaged) over
final (initial) state spin and color states is
\begin{equation}
\label{eq:born_t}
\overline{\sum} \left|{\cal M}_{t,\rm Born}\right|^2 = 
\frac{1}{4} g^4 |V_{ud}|^2 |V_{tb}|^2 s(s-m_t^2)
\left| \frac{1}{t-M^2_{W}}\right|^2 \, .
\end{equation}
Here, $s=(p_u + p_b)^2$ and $t=(p_u-p_d)^2$, the partonic reaction
sub-energy squared and the square of the momentum transfer across the
$W$, respectively.  The CKM matrix elements $|V_{ij}|$ may be changed
for the given particles, and $m_t$ is the top-quark mass.  The result
for the $s$-channel is obtained by interchanging $s$ and $t$, and
letting $s=(p_u + p_{\bar d})^2$ and $t=(p_u-p_{\bar b})^2$.  In Sec.\
\ref{sec:dipole-subtraction} we discuss the cross section without the
sum over spins.

\begin{figure}[hbtp]
\begin{center}
\centerline{\epsfig{figure=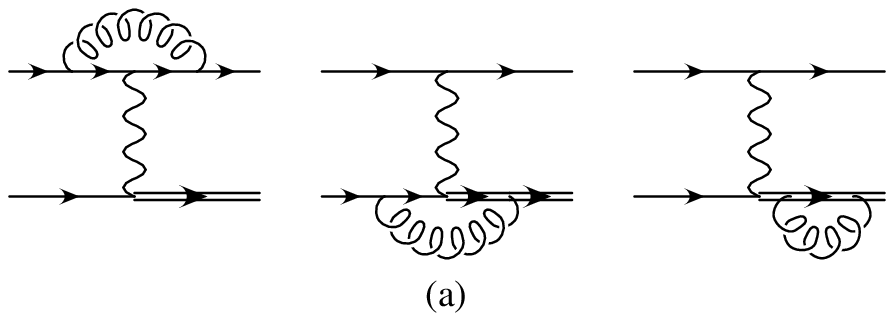,width=3.5in}}
\centerline{\epsfig{figure=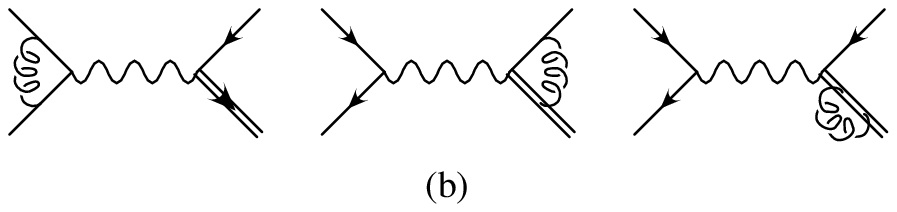,width=3.5in}}
\caption{\small The one-loop virtual corrections to the (a)
$t$-channel and (b) $s$-channel amplitudes.\label{fig:2}}
\end{center}
\end{figure}

At next-to-leading order we must include the virtual QCD corrections
to Eqs.\ (\ref{eq:3}) and (\ref{eq:6}), shown in Fig.~\ref{fig:2}.
To obtain the real emission contributions one must attach a gluon
in all possible ways to the diagrams in Fig.~\ref{fig:1}. 
The resulting crossed diagrams may be written as
\begin{eqnarray}
u + \bar{d} &\rightarrow & t + \bar{b} + g \,, \label{eq:real1} \\
u + b &\rightarrow & t + d + g \nonumber \\
\bar{d} + b &\rightarrow & t + \bar{u} + g \,, \label{eq:real2} \\
u + g &\rightarrow & t + \bar{b} + d \nonumber \\
\bar{d} + g &\rightarrow & t + \bar{b} + \bar{u} \,, \label{eq:real3} \\
g + b &\rightarrow & t + d + \bar{u} \,. \label{eq:real4}
\end{eqnarray}
These crossings may belong either to the $s$ or $t$ channel, and their
assignment is governed by the collinear singularity structure.  It is
useful to distinguish whether the gluon attaches to the fermion
current containing the heavy-quark line ($h$) in Fig.~\ref{fig:1} or
to the current containing only the light-quark lines ($l$).  Note that
the contributions from $h$ and $l$ radiation do not interfere in the
cross section due to color
conservation~\cite{Smith:1996ij,Stelzer:1997ns}.

The diagrams in Eq.~(\ref{eq:real1}) belong fully to the $s$-channel,
and diagrams in Eq.~(\ref{eq:real2}) belong to the $t$-channel.  For
crossing in Eq.~(\ref{eq:real3}) let us consider the first process $u
+ g \rightarrow t + \bar{b} + d$.  In the heavy-quark line the gluon
can split into a collinear $b\bar{b}$-pair with reduced process $u + b
\rightarrow t+ d$, so that this crossing belongs to the
$t$-channel. In the light-quark line the gluon can split into a
collinear $d\bar{d}$-pair, with reduced process $u + \bar{d}
\rightarrow t + \bar{b}$, so that this crossing belongs to the
$s$-channel.  The classification of the second process in
Eq.~(\ref{eq:real3}) is similar, with the role of the $u$- and
$d$-type quarks exchanged.  Finally, in Eq.~(\ref{eq:real4}), the
gluon connected to the light-quark line can split either into a
collinear $u\bar{u}$-pair or into a collinear $d\bar{d}$-pair.
In both cases the process reduces to the $t$-channel process $u + b
\rightarrow t + d$ or $\bar{d} + b \rightarrow t + \bar{u}$.  Crossing
of Eq.~(\ref{eq:real4}) to the heavy-quark line is not included in our
calculation as it involves on-shell $W$-decay into light fermions, and
is classified as $W$-top associated production. The cross section for
this process is estimated to be negligibly small for the Tevatron
\cite{Heinson:1996zm,Stelzer:1998ni,Belyaev:1998dn,Tait:1999cf,Belyaev:2000me}.

In what follows we consider only the production of a single top quark,
but the relevant expressions for the production of an antitop quark
may easily be obtained by charge conjugation.  Our calculational
framework is further specified as follows: we set the bottom quark
mass to zero, so that we work with 5 massless flavors in the parton
distribution functions.  It is important to use a consistent set of
5-flavor parton distribution functions, in which the 5-flavor set has
been computed from a lower flavor number set via NLO matching
conditions \cite{Collins:1986mp} that preserve the momentum sum rule.
To handle divergences occurring at intermediate stages of the
calculation we use dimensional regularization ($D=4-2\epsilon$).
Details describing renormalization of
the vertices that contain $\gamma_5$ are given in
Sec.~\ref{sec:scheme}.  Collinear divergences are subtracted in the
standard $\overline{\mathrm{MS}}$ scheme. We use the Feynman gauge for
the gluon propagator.

\section{Phase space slicing}
\label{sec:slicing}

In the phase space slicing method a subregion of phase space
containing soft and collinear singularities is defined.  If the
subregion is small enough, one may make simplifying kinematic
approximations in the phase space integrals, so that they can be
performed analytically.  To define the size and shape of this
subregion one introduces one or two theoretical cutoff parameters
(for common or separate treatment of collinear and soft contributions,
respectively).  The final cross section should not depend on the
choice made for the partitioning.  Below we discuss the application of
these methods to single-top-quark production.

\subsection{Phase space slicing with two cutoffs ($\delta_s,\delta_c$)}
\label{sec:phase-space-slicing}

A detailed description of the phase space slicing method with two
cutoffs has recently been given in Ref.\ \cite{Harris:2001sx}.  In
this section we present the results needed for single-top-quark
production.  We follow closely the notation of Ref.\
\cite{Harris:2001sx}.

In the two cutoff method, phase space is divided into a hard (H) and
soft (S) region.  The contribution from the latter region is computed
by applying the eikonal approximation to the radiative processes in
Eqs.~(\ref{eq:real1}--\ref{eq:real4}).  The soft region of phase space
is defined by a condition on the energy of the soft gluon in the
partonic center-of-momentum frame:
\begin{equation}
  \label{eq:15}
  0 \leq E_g \leq \delta_s\, \frac{\sqrt{s}}{2}\,.
\end{equation}
The hard region is then defined by $E_g > \delta_s \sqrt{s}/2$.  After
performing the $D$-dimensional angular integrations of the gluon in
the region defined in Eq.\ (\ref{eq:15}), the 
soft contribution is found to be
\begin{equation}
  \label{eq:16}
  d\sigma_p^{(S)} =   d\sigma_p^{(0)} 
  \left[
    \frac{\alpha_s}{2\pi}\frac{\Gamma(1-\epsilon)}{\Gamma(1-2\epsilon)}
    \left(
      \frac{4\pi\mu^2}{s}
    \right)^\epsilon
  \right]\,
  \left(
    \frac{A_2^p}{\epsilon^2}+ \frac{A_1^p}{\epsilon}+A_0^p
  \right) \,,
\end{equation}
with $p=s,t$ labeling the channel.  For the $s$-channel 
\begin{eqnarray}
  \label{eq:17}
  A_2^s &=& 3C_F \nonumber \\
  A_1^s &=& C_F
  \left[
    1-6\ln\delta_s -\ln\left(\frac{s}{m^2}\right)
  \right]\nonumber \\
  A_0^s &=& C_F
  \Bigg[
    6\ln^2\delta_s -2\ln\delta_s +
  2\ln\delta_s\ln\left(\frac{s}{m^2}\right) \nonumber \\
 & + & \frac{s+m^2}{s-m^2}\ln\left(\frac{s}{m^2}\right) -2
  \mathrm{Li}_2(\beta)
  -\frac{1}{2}\ln^2\left(\frac{s}{m^2}\right)
  \Bigg] \,,
\end{eqnarray}
and for the $t$-channel
\begin{eqnarray}
\label{eq:18}
  A_2^t &=& 3C_F \nonumber \\
  A_1^t &=& C_F
  \left[
    1-6\ln\delta_s -2\ln\left(\frac{-t}{s\beta}\right)
    -\ln\left(\frac{(m^2-t)^2}{m^2s}\right)
  \right]\nonumber \\
  A_0^t &=& C_F
  \Bigg[
    6\ln^2\delta_s -2\ln\delta_s +
  4\ln\delta_s\ln\left(\frac{-t}{s\beta}\right) \nonumber \\
 &\ & +2\ln\delta_s\ln\left(\frac{(m^2-t)^2}{m^2s}\right) +
\frac{s+m^2}{s-m^2}\ln\left(\frac{s}{m^2}\right) 
 +  \ln^2\left(\frac{-t}{s\beta}\right)
  +2\mathrm{Li}_2\left(1+\frac{t}{s\beta}\right)\nonumber \\
&\ &  -\frac{1}{2}\ln^2\left(\frac{s}{m^2}\right)+\ln^2\left(\frac{m^2}{m^2-t} \right)
+2\mathrm{Li}_2\left(\frac{t}{m^2}\right)
-2\mathrm{Li}_2\left(\frac{u}{s+u}\right)
  \Bigg] \,,
\end{eqnarray}
where the top-quark mass is denoted as $m$, and $\beta=1-m^2/s$.

The hard region is further divided into hard collinear (HC) and
hard-noncollinear (H$\overline{\mathrm{C}}$) regions. The latter is
computed in 4 dimensions, integrating numerically over the
H$\overline{\mathrm{C}}$ phase space using standard Monte Carlo
methods.  The HC contributions arise from integration over those
regions of phase space where any invariant, $s_{ij}=(p_i+p_j)^2$ or
$t_{ij}=(p_i-p_j)^2$, appearing in the denominator becomes smaller in
magnitude than $\delta_c s$.  The singular regions are distinguished
by whether they come from initial or final state radiation.  The
latter are given by
\begin{equation}
  \label{eq:20}
    d\sigma_p^{(HC,FS)} =   d\sigma_p^{(0)} 
  \left[
    \frac{\alpha_s}{2\pi}\frac{\Gamma(1-\epsilon)}{\Gamma(1-2\epsilon)}
    \left(
      \frac{4\pi\mu^2}{s}
    \right)^\epsilon
  \right]\,
  \left(
    \frac{A_1}{\epsilon}+ A_0
  \right) \,,
\end{equation}
with
\begin{eqnarray}
  \label{eq:21}
    A_1 &=& C_F
    \left(
      2\ln \delta_s+\frac{3}{2}-2\ln\beta
    \right)
\nonumber \\
  A_0 &=& C_F
  \Bigg[
\frac{7}{2}-\frac{\pi^2}{3}-\ln^2\delta_s-\ln^2\beta + 2\ln\delta_s
    \ln\beta \nonumber \\ 
&& -\ln\delta_c
\left(
  2\ln\delta_s + \frac{3}{2}-2\ln\beta
\right)
  \Bigg] \,,
\end{eqnarray}
for both channels (all massless partons in the Born cross sections are
fermions).

The collinear contributions from the initial state are given by the 
sum of two contributions.  The first is the finite remainder after mass 
factorization.  The second results from a mismatch in integration limits 
when subtracting the mass factorization counter-term.
\begin{equation}
  \label{eq:22}
  d\sigma_{p,C}^{ij\rightarrow FS} = d\sigma_p^{(0)}
  \left[
    \frac{\alpha_s}{2\pi}\frac{\Gamma(1-\epsilon)}{\Gamma(1-2\epsilon)}
    \left( \frac{4\pi\mu^2}{s} \right)^\epsilon \right] \left[
 \widetilde{f}_j^H(z,\mu_F)
 +\left( \frac{A_1^{sc}}{\epsilon}+A_0^{sc}\right)
  f_j^H \left(z,\mu_F\right) \right] \,.
\end{equation}
The modified parton distribution function $\widetilde{f}$ is 
given in Ref.\ \cite{Harris:2001sx}.
For the reaction at hand we only need consider the quark-quark
initial state splitting, so
\begin{eqnarray}
\label{eq:26}
  A_1^{sc} &=& C_F\left(2\ln\delta_s + \frac{3}{2} \right)\nonumber\\
  A_0^{sc} &=& C_F\left(2\ln\delta_s + \frac{3}{2} \right)\,
  \ln\left(\frac{s}{\mu_F^2}   \right) \; .
\end{eqnarray}
The virtual contribution is obtained as explained in the
Secs.~\ref{sec:calc} and \ref{sec:scheme}.  The results in the
notation of this section are
\begin{equation}
  \label{eq:101}
  d\sigma_p^{(V)}= d\sigma_p^{(0)} 
  \left[
    \frac{\alpha_s}{2\pi}\frac{\Gamma(1-\epsilon)}{\Gamma(1-2\epsilon)}
    \left(
      \frac{4\pi\mu^2}{s}
    \right)^\epsilon
  \right]\,
  \left(
    \frac{A_2^V}{\epsilon^2}+ \frac{A_1^V}{\epsilon}+A_0^V
  \right) +  \left( \frac{\alpha_s}{2\pi} \right) d\tilde{\sigma}_p^{(V)} \,,
\end{equation}
where
\begin{eqnarray}
  \label{eq:102}
  A_2^V &=& C_F \left\{ \left[ -2 \right] - \left[ 1 \right] \right\} \nonumber \\
  A_1^V &=& C_F \left\{ 
  \left[ -3 -2 \ln\left( \frac{s}{-q^2} \right) \right] + \left[ -\frac{5}{2} 
   - 2 \ln \left( 1 - \lambda \right) - \ln\left( \frac{s}{m^2} \right) 
\right] \right\} \nonumber \\
  A_0^V &=& C_F \left\{ \left[ -\ln^2 \left( \frac{s}{-q^2} \right) 
- 3 \ln \left( \frac{s}{-q^2} \right) -8 - \frac{\pi^2}{3} \right] 
\right. \nonumber \\
&+& \left. \left[ -\frac{1}{2}\ln^2\left(\frac{s}{m^2}\right)
-\frac{5}{2}\ln\left(\frac{s}{m^2}\right) - 2\ln\left( 1 - \lambda \right)
\ln\left(\frac{s}{m^2}\right) - 6 \right. \right. \nonumber \\ 
&-& \left. \left. \frac{1}{\lambda}\ln\left( 1 - \lambda \right)-\ln^2\left( 1 - \lambda \right)-2\ln\left( 1 - \lambda \right)
+2{\rm Li}_2\left( \lambda \right) - \frac{\pi^2}{3} 
\right] \right\} \; .
\end{eqnarray}
In the above we have defined $\lambda\equiv q^2/(q^2-m^2)$.  Further, the 
separate terms in the square brackets originate from the massless-massless or 
the massive-massless vertex corrections.  For the $t$-channel one sets 
$q^2=t$, while for the $s$-channel one sets $q^2=s$.
\begin{equation}
d\tilde{\sigma}_t^{(V)} = \frac{1}{2s} \frac{1}{4} g^4 |V_{ud}|^2 
|V_{tb}|^2 C_F \frac{m^2su}{t}\ln\left(\frac{m^2}{m^2-t} 
\right) \left( \frac{1}{t-M_W^2} \right)^2 d\Gamma_2 \,,
\end{equation}
is the $t$-channel finite piece in the virtual contribution that is not 
proportional to the Born cross section.  It results from the 
interference of the renormalized massive-massless vertex 
with the Born amplitude.  The $s$-channel version may be obtained by crossing.

At this point one can see that the two body weight is finite:
$A_2^p+A_2^V=0$ and $A_1^p+A_1^V+A_1+2A_1^{sc}=0$.
The factor of two occurs since there are two quark legs, either 
of which can emit a gluon.  The final finite two-body cross 
section is given by the sum of the residual $\widetilde{f}$ terms
from both the quark-quark and quark-gluon initiated processes 
and the finite two-body weights.  The result, summed over parton flavors is
\begin{eqnarray}
\sigma^{(2)} &=&\left( \frac{\alpha_s}{2\pi} \right)
\sum_{a,b} \int dx_1 dx_2 \left\{  f_a^{H_1}(x_1,\mu_F) 
f_b^{H_2}(x_2,\mu_F) \left[ d\sigma^{(0)}_p 
\left( A_0^p+A_0^V+A_0+2A_0^{sc} \right) + d\tilde{\sigma}_p^{(V)}
\right] \right. \nonumber \\
&+& \left. d\sigma^{(0)}_p \left[ f_a^{H_1}(x_1,\mu_F) 
\widetilde{f}_b^{H_2}(x_2,\mu_F) 
+ \widetilde{f}_a^{H_1}(x_1,\mu_F) f_b^{H_2}(x_2,\mu_F) \right] 
+ (x_1 \leftrightarrow x_2)\right\} \, .
\end{eqnarray}

\newcommand{\sot}{s^{}_{12}}
\newcommand{\stfp}{s^{\prime}_{34}}
\newcommand{\stf}{s^{}_{35}}
\newcommand{\sffp}{s^{\prime}_{45}}
\newcommand{\tot}{t^{}_{13}}
\newcommand{\ttt}{t^{}_{23}}
\newcommand{\tofp}{t^{\prime}_{14}}
\newcommand{\ttfp}{t^{\prime}_{24}}
\newcommand{\tof}{t^{}_{15}}
\newcommand{\ttf}{t^{}_{25}}

The three-body contribution is given by
\begin{equation}
\sigma^{(3)} = \sum_{a,b} \int dx_1 dx_2 \frac{1}{2s}\int_{H\overline C}
\overline{\sum} |M_3^{(ab)}|^2 d\Gamma_3 \,,
\end{equation}
with
\begin{equation}
\overline{\sum} |M_3^{(ab)}|^2 = -\pi g^4 |V_{ud}|^2 |V_{tb}|^2
\Psi_i \;\;, i=1-3 \,.
\end{equation}

The $\Psi_i$ contain the luminosity and Dirac algebra of
Eqs.~(\ref{eq:real2}--\ref{eq:real4}).  We choose initial-state
momenta as incoming, and label the momenta for the $t$-channel as
\begin{eqnarray}
\Psi_1: u(p_1) b(p_2) &\to& d(p_3) t(p_4) g(p_5) \,,\\
\Psi_2: u(p_1) g(p_2) &\to& d(p_3) t(p_4) \bar b(p_5) \,,\\
\Psi_3: g(p_1) b(p_2) &\to& d(p_3) t(p_4) \bar u(p_5) \,.
\end{eqnarray}
The $t$-channel $\Psi_i$ are given by
\begin{eqnarray}
\Psi_1&=& 2 C_F 
F_1 \left( \frac{\sot\left( \tofp-\stfp\right) -\stfp\ttf}{\tof} -
\frac{\sot \left( \stfp+\sffp\right) - \stfp\ttt}{\stf} \right. \nonumber \\
&&\quad\quad\quad\quad\left. -\frac{\tot\left[ \sot \left( 2\stfp+
\sffp\right) + \stfp\ttf\right]}{\tof\stf} \right) \nonumber \\
&&+ 2 C_F F_2 \left( \frac{\sot\left( \ttt-\stfp\right) -\stfp\tof}{\ttf} -
\frac{\sot\left( \stfp+\stf \right) \left( 1-2m_t^2/\sffp\right)
- \stfp\tofp}{\sffp} \right. \nonumber \\
&&\quad\quad\quad\quad\quad\left. -\frac{\ttfp\left[ \sot \left( 2\stfp+
\stf\right) + \stfp\tof\right]}{\ttf\sffp} \right)  \;, \\
\Psi_2&=& F_2 \left( \frac{\sot\stfp + \tof\left( \stfp-\stf\right)}{\ttf}
+\frac{\tof\left( \stfp+\ttt\right) \left( 1- 2m_t^2/\ttfp\right) - \stfp
\tofp}{\ttfp} \right. \nonumber \\
&&\quad\quad\left. + \frac{\sffp\left[ \sot\stfp + \tof\left( \ttt + 2\stfp
\right) \right]}{\ttfp\ttf} \right)
\;, \\
\Psi_3&=& F_1 \left( \frac{\sot\stfp + \ttf\left( \stfp-\sffp\right)}{\tof}
+\frac{\ttf\left( \stfp+\tofp\right) - \stfp\ttt}{\tot} \right. \nonumber \\
&&\quad\quad\left. + \frac{\stf\left[ \sot\stfp + \ttf\left( \tofp + 2\stfp
\right) \right]}{\tot\tof} \right) \;,
\end{eqnarray}
where $C_F=4/3$, $s_{ij}=(p_i+p_j)^2$, $t_{ij}=(p_i-p_j)^2$,
$s_{ij}^\prime = s_{ij}-m_t^2$, $t_{ij}^\prime = t_{ij}-m_t^2$,
\begin{eqnarray}
F_1=\frac{\alpha_{s\,l}}{\left(t^{}_{24}-M_W^2\right)^2}L_l \;,\\
F_2=\frac{\alpha_{s\,h}}{\left(\tot-M_W^2\right)^2}L_h \;,
\end{eqnarray}
$\alpha_{s\,l(h)}$ and the luminosity functions $L_{l(h)} =
f_a^{H_1}(x_1,\mu_{F\,l(h)})f_b^{H_2}(x_2,\mu_{F\,l(h)})$ are
evaluated using the scales at the light(heavy)-quark lines,
respectively.  All other $s$- and $t$-channel matrix elements can be
obtained by crossing.  Physical predictions follow from the sum
$\sigma^{(2)}+\sigma^{(3)}$, which is cutoff independent for
sufficiently small cutoffs as shown below.

\subsection{Phase space slicing with one cutoff ($\smin$)}
\label{sec:phase-space-slicing-1}

The calculation using the one cutoff slicing method is similar to the
one using the two cutoff slicing method, with some differences that we
now address.  In this method, a pair of partons with momenta $p_i$ and
$p_j$ is defined to be unresolved if
\begin{equation}
  \label{eq:1}
  |2 p_i\cdot p_j| < \smin \,,
\end{equation}
with $\smin$ small compared to the hard scale of the process.  The
condition in Eq.\ (\ref{eq:1}) can occur if either $p_i$ and $p_j$ are
collinear, or if one of the two is soft.  This method, combined with
the use of color-decomposed amplitudes and universal crossing
functions, has been developed into a general method for computing with
minimal calculational effort fully differential NLO production cross
sections of bosons and jets in \cite{Giele:1992vf,Giele:1993dj} and
identified hadrons and heavy quarks in \cite{Keller:1998tf}.  The
single-top-quark production process has a relatively simple color
structure, so we do not need to decompose the scattering amplitudes
into color-ordered subamplitudes.

The treatment of the virtual corrections is no different from the two
cutoff method.  To determine the radiative corrections, all partons
are first crossed to the final state, and resolved and unresolved
contributions are identified according to the criterion in Eq.\
(\ref{eq:1}). The unresolved contributions, (soft and collinear) can
be found in \cite{Giele:1992vf,Giele:1993dj,Keller:1998tf} expressed
in $D=4-2\epsilon$ dimensions.  The soft contributions are expressed
in terms of the momenta for partons in lowest order kinematics with
all partons in the final state
\begin{equation}
  \label{eq:7}
  0\rightarrow \bar{u} + d + t + \bar{b}\,,
\end{equation}
and are given by
\begin{equation}
  \label{eq:9}
  d\sigma^{(S)} = d\sigma^{(0)} \left[ \frac{\alpha_s\,
  C_F}{\pi}\frac{1}{\Gamma(1-\epsilon)}
\left(\frac{4\pi\mu^2}{\smin}\right)^\epsilon \right]
\left[\frac{1}{\epsilon^2}\left(\frac{2p_u\cdot p_d}{\smin}\right)^\epsilon
 + {\cal J}(m,0)\left(\frac{2p_t\cdot p_b}{\smin}\right)^\epsilon\right] \,,
\end{equation}
with $ d\sigma^{(0)}$ obtained by crossing all momenta to the final
state, and
\begin{eqnarray}
  \label{eq:12}
  1)\;\; 2p_t\cdot p_b \geq m^2 \;:\;\;\;
 {\cal J}(m,0) &=& \frac{1}{\epsilon^2} 
- \frac{1}{2\epsilon^2}\left( \frac{2p_t\cdot p_b}{m^2}\right)^\epsilon
+\frac{1}{2\epsilon} \left( \frac{2p_t\cdot p_b}{m^2} \right)^\epsilon
-\frac{\pi^2}{12} + \frac{m^2}{2p_t\cdot p_b} \nonumber \\
2)\;\; 2p_t\cdot p_b \leq m^2 \;:\;\;\;
{\cal J}(m,0)&=& \left( \frac{2p_t\cdot p_b}{m^2} \right)^{-\epsilon}
\left( \frac{1}{2 \epsilon^2} 
+\frac{1}{2\epsilon} 
-\frac{\pi^2}{12} + 1 \right) \,.
\end{eqnarray}
The $s$-channel contribution is then obtained by replacing
$2p_u\cdot p_d \rightarrow s$, $2p_t\cdot p_b\rightarrow s-m^2$.
The $t$-channel contribution is obtained by replacing
$2p_u\cdot p_d \rightarrow t$, $2p_t\cdot p_b\rightarrow t-m^2$,
which leads to $\pi^2$ terms after expanding in $\epsilon$.
The collinear contributions are likewise given by
\begin{eqnarray}
  \label{eq:10}
  d\sigma^{(C)} &=& - d\sigma^{(0)} \left[ \frac{\alpha_s\,
  C_F}{\pi}\frac{1}{\Gamma(1-\epsilon)}
\left(\frac{4\pi\mu^2}{\smin}\right)^\epsilon \right] \frac{1}{\epsilon}
\nonumber\\
&&\times \left[I_{q\rightarrow qg}\left(0,\frac{\smin}{2p_u\cdot p_d}\right)
+ I_{\bar{q}\rightarrow {q}g}\left(\frac{\smin}{2p_u\cdot p_d},0\right)
+ I_{\bar{q}\rightarrow {q}g}\left(\frac{\smin}{2p_t\cdot p_b},0\right)
\right] \,.
\end{eqnarray}
The $I$ functions are given in Refs.\ \cite{Giele:1992vf,Giele:1993dj}.
The $s$-channel contribution is obtained by replacing $2p_u\cdot p_d
\rightarrow s$, $2p_t\cdot p_b\rightarrow s-m^2$.  The $t$-channel
contribution is obtained by replacing $2p_u\cdot p_d \rightarrow t$,
$2p_t\cdot p_b\rightarrow t-m^2$.  The sum of these contributions is
already finite. One now generates the various subprocesses of
single-top-quark production by crossing pairs of partons back to the
initial state.  Crossing symmetry is not a property of next-to-leading
order cross sections, but it may be implemented in the following way.

In general a NLO fully differential cross section for a process with initial
hadrons $H_1$ and $H_2$ may be written as
\begin{equation}
  \label{eq:one}
  d\sigma_{H_1\,H_2} = \sum_{a,b}\int dx_1\int dx_2
\, {\cal F}_a^{H_1}(x_1) \,{\cal F}_b^{H_2}(x_2) d\sigma_{ab}^{NLO}(x_1,x_2)\ ,
\end{equation}
where $a,b$ denote parton flavors and $x_1,x_2$ are parton momentum fractions.
The function $d\sigma_{ab}^{NLO}$ is computed with 
all-partons-in-the-final-state matrix elements,
in which partons $a$ and $b$ have simply been crossed to the initial state,
i.e.\ in which their momenta $p_a$ and $p_b$ have been replaced by 
$-p_a$ and $-p_b$ (this function does include the
$\pi^2$ terms resulting from this replacement in the one-loop
virtual graphs). The functions ${\cal F}_a^{H}(x)$ are 
modifications of the parton distribution functions $f_a^{H}(x,\mu_F)$
\begin{equation}
  \label{eq:two}
  {\cal F}_a^{H}(x) = f_a^{H}(x,\mu_F) + \alpha_s C_a^{H}(x,\mu_F)
 + O(\alpha_s^2),
\end{equation}
where $C_a^{H}(x,\mu_F)$ are finite, universal ``crossing functions'' \cite{Giele:1993dj}.
They implement the crossing property for the unresolved contributions, and 
are given by
\begin{equation}
  \label{eq:13}
  C_a^{H,\msb}(x,\mu_F) = \frac{N_C}{2\pi}\left[A_a^H(x,\mu_F)
 \ln\left(\frac{\smin}{\mu_F^2}\right)
+B_a^{H,\msb}(x,\mu_F)\right]\, .
\end{equation}
The functions $A_a^H,B_a^H$ functions for the proton are given
in \cite{Giele:1993dj}. 
In the unresolved contribution one may simply cross pairs of partons
without further modifications. 

The full NLO differential cross section can now be written as:
\begin{eqnarray} \label{eq:11}
d\sigma_{H_1\, H_2} &= &\sum_{a,b}\int dx_1 dx_2
 f_a^{H_1}(x_1,\mu_F) \,f_b^{H_2}(x_2,\mu_F) 
  \, \Big[ d\sigma_{ab}^{NLO}(x_1,x_2) + \alpha_s (\mu_F) \nonumber \\
&&\times\Big( C^{H_1}_a(x_1,\mu_F) f_b^{H_2}(x_2,\mu_F) +
f_a^{H_1}(x_1,\mu_F) C^{H_2}_b(x_2,\mu_F) \Big)
d\sigma_{ab}^{LO}(x_1,x_2) \Big] \,.
\end{eqnarray}
The unresolved contribution, now including the crossing functions,
depends analytically on $\smin$, but this $\smin$ dependence cancels
against that of the resolved contribution.  The results produced with
this method agree with those of the previous section. In this paper we
limit ourselves to some illustrative numerical studies, so that we
only employ the dipole and two-cutoff slicing methods for numerical
results.

\section{Massive Dipole Subtraction Calculation}
\label{sec:dipole-subtraction}

Within the dipole formalism the 
NLO cross section is rewritten as
\begin{eqnarray}
\sigma^{NLO} & = & \int\limits_{n+1} d\sigma^R + \int\limits_n d\sigma^V
\nonumber \\
& = & \int\limits_{n+1} \left( d\sigma^R - d\sigma^A \right) + \int\limits_n
\left( d\sigma^V + \int\limits_1 
d\sigma^A \right)\,.
\end{eqnarray}
In the second line an approximation term $d\sigma^A$ has been added
and subtracted.  This is valid if all singularities occur in the final
state.  For initial state partons there are slight modifications.  The
approximation $d\sigma^A$ has to fulfill the requirement that
$d\sigma^A$ is a proper approximation of $d\sigma^R$ such as to have
the same point-wise singular behavior (in $D$ dimensions) as
$d\sigma^R$ itself.  Thus, $d\sigma^A$ acts as a local counter-term for
$d\sigma^R$ and one can safely perform the limit $\varepsilon
\rightarrow 0$. This defines a cross-section contribution
\begin{eqnarray}
\sigma^{NLO}_{\{n+1\}} & = & \int\limits_{n+1} \left( \left. d\sigma^R
\right|_{\varepsilon=0} - \left. d\sigma^A \right|_{\varepsilon=0} \right)\,.
\end{eqnarray}
$d\sigma^A$ is analytically integrable (in $D$ dimensions) over the
one-parton subspace leading to soft and collinear divergences. This
gives the contribution
\begin{eqnarray}
\sigma^{NLO}_{\{n\}} & = & \int\limits_n \left( d\sigma^V + \int\limits_1
d\sigma^A \right)_{\varepsilon=0}\,.
\end{eqnarray}
The final structure of an NLO calculation is then
\begin{eqnarray}
\label{NLOcrosssection}
\sigma^{NLO} & = & \sigma^{NLO}_{\{n+1\}} + \sigma^{NLO}_{\{n\}}.
\end{eqnarray}

Since both contributions on the right hand side of
Eq.\ (\ref{NLOcrosssection}) are now finite, they can be evaluated with
numerical methods.  The $(n+1)$ matrix element is approximated by a
sum of dipole terms
\begin{eqnarray}
\lefteqn{d\sigma^A \sim \sum\limits_{\mathrm{pairs}\; i,j} \sum\limits_{k
\neq i,j} {\cal D}_{ij,k} } \nonumber \\
& = & 
\sum\limits_{\mathrm{pairs}\; i,j} \sum\limits_{k \neq i,j} 
- \frac{1}{2 p_i \cdot p_j}
\langle 1, ..., \tilde{(ij)},...,\tilde{k},...|
\frac{{\bf T}_k \cdot {\bf T}_{ij}}{{\bf T}^2_{ij}} V_{ij,k} |
1,...,\tilde{(ij)},...,\tilde{k},... \rangle\,, \nonumber \\
\end{eqnarray}
where the emitter parton is denoted by $\tilde{ij}$ and the spectator
by $\tilde{k}$.  Here ${\bf T}_i$ denotes the color charge operator
\cite{Catani:1997vz} for parton $i$ and $V_{ij,k}$ is a matrix in the
helicity space of the emitter with the correct soft and collinear
behavior.  $|1,...,\tilde{(ij)},...,\tilde{k},... \rangle$ is a vector
in color- and helicity space.  By subtracting from the real emission
part the fake contribution we obtain
\begin{eqnarray}
d \sigma^R - d \sigma^A & = & d\phi_{n+1} 
\Biggl( |M(p_1,...,p_{n+1})|^2 \theta^{cut}_{n+1}(p_1,...,p_{n+1})
\nonumber \\
& &  - \sum\limits_{\mathrm{pairs}\; i,j} \sum\limits_{k \neq i,j}
{\cal D}_{ij,k}(p_1,...,p_{n+1}) \theta^{cut}_{n}(p_1,...,\tilde{p}_{ij},...,
\tilde{p}_k,...,p_{n+1}) \Biggr) \,. \nonumber \\
\end{eqnarray}
Both $d\sigma^R$ and $d\sigma^A$ are integrated over the same $(n+1)$
parton phase space, but it should be noted that $d\sigma^R$ is
proportional to $\theta^{cut}_{n+1}$, whereas $d\sigma^A$ is
proportional to $\theta^{cut}_n$.  Here $\theta^{cut}_n$ denotes the
jet-defining function for $n$-partons.

The subtraction term can be integrated over the one-parton phase space
to yield the term
\begin{eqnarray}
{\bf I} \otimes d\sigma^B & = & \int\limits_{1} d\sigma^A =
\sum\limits_{\mathrm{pairs}\; i,j} \sum\limits_{k \neq i,j} \int
d\phi_{dipole} {\cal D}_{ij,k}\,.
\end{eqnarray}
The universal factor ${\bf I}$ still contains color correlations, but
does not depend on the unresolved parton $j$.  The term ${\bf I}
\otimes d\sigma^B$ lives on the phase space of the $n$-parton
configuration and has the appropriate singularity structure to cancel
the infrared divergences coming from the one-loop amplitude.
Therefore,
\begin{equation}
d\sigma^V + {\bf I} \otimes d\sigma^B
\end{equation}
is infrared finite and can easily be integrated by Monte Carlo
methods.  The explicit forms of the dipole terms ${\cal D}_{ij,k}$,
together with the integrated counterparts, can be found in Ref.\
\cite{Catani:1997vz} (the original massless case) and in
Ref.\ \cite{Phaf:2001gc} (extension to massive fermions).

\subsection{Calculation of the amplitudes}
\label{sec:calc}

We have performed three different calculations of the loop
amplitudes.  One calculation was done using the standard approach in
the 't~Hooft-Veltman scheme.  The second calculation involved treating
the $\gamma_5$ as anti-commuting in $D$ dimensions, thereby retaining
Ward identities for the charged current vertex.  In the third one we
calculated helicity amplitudes using a four-dimensional scheme.  The
effects of different prescriptions for $\gamma_5$ are discussed more
extensively in Sec.\ \ref{sec:scheme}.  The results of the three
calculations agree with each other, in the sense that they can be
related to each other through process-independent finite
renormalizations.  In addition, we find agreement with the earlier
calculations of Refs.~\cite{Gottschalk:1981rv,Gluck:1996ve}.  We
present here the helicity amplitudes obtained with the third approach.
They are more compact and contain the complete spin information.  In
the standard approach one just calculates the interference between the
loop amplitude and the Born term and sums over all spins.

We first list our conventions (for reviews of spinor helicity methods
see e.g.~\cite{Mangano:1991by,Dixon:1996wi}).  With spinor helicity
methods we can express scattering amplitudes in terms of massless Weyl
spinors of helicity $\pm\frac{1}{2}$,
\begin{eqnarray}
\label{eq:19}
u(p,\pm) = v(p,\mp) =  | p \pm \r \,, \qquad   \bar{u}(p,\pm) = \bar{v}(p,\mp) = \l p \pm |  \,.
\end{eqnarray}
External fermion states are directly expressed in terms of these.  Our
convention is to take all particles outgoing.  For example, an
outgoing massless fermion with positive helicity is denoted by $\l p +
|$, while an outgoing massless anti-fermion with positive helicity is
denoted by $| p - \r$.  The gluon polarization vectors, of helicity
$\pm 1$, may be written as
\begin{eqnarray} \label{eq:23}
\varepsilon^+_\mu(k,q) = \frac{\l q- | \g_\mu | k- \r}{\sqrt{2}\l qk \r}, & &
\varepsilon^-_\mu(k,q) = \frac{\l q+ | \g_\mu | k+ \r}{\sqrt{2} [ kq ]}.
\end{eqnarray}
We use the customary short-hand notation:
\begin{eqnarray}\label{eq:24}
\l ij \r = \l p_i - | p_j + \r, & & [ ij ] = \l p_i + | p_j - \r .
\end{eqnarray}
In Eq.\ (\ref{eq:23}) $k$ is the gluon momentum and $q$ an arbitrary
light-like ``reference momentum''.  The dependence on the choice of
$q$ drops out in gauge-invariant amplitudes.  We shall also employ the
abbreviations
\begin{eqnarray}\label{eq:25}
\l i- | k+l | j- \r & = & \l ik \r [ kj] + \l il \r [ lj ], \nonumber \\
s_{ij...k} & = & \left( p_i + p_j + ... + p_k \right)^2  ,
\end{eqnarray}
with all momenta null-vectors.

To treat the massive top quark within the framework of spinor helicity
methods, we use the extension to massive fermions
\cite{Kleiss:1985yh,Berends:1985gf,Ballestrero:1995jn,Dittmaier:1998nn}.
Even though helicity is not a conserved quantum number for a massive
particle, a massive positive-energy spinor satisfying the Dirac
equation has a two-fold degeneracy (labeled by a spin-component
quantized along some axis).  With slight abuse of notation we label
these two states by ``$+$'' and ``$-$''.  Let $p$ be a four-vector
with $p^2=m^2$ and $p_0 > 0$, and let $q$ be an arbitrary null vector
with $q_0 > 0$.  We define
\begin{eqnarray}\label{eq:4}
u(p,+) = \frac{1}{\sqrt{2 p q}} \left( p\!\!\!/ + m \right) | q - \r,
& & 
v(p,+) = \frac{1}{\sqrt{2 p q}} \left( p\!\!\!/ - m \right) | q - \r,
\nonumber \\
u(p,-) = \frac{1}{\sqrt{2 p q}} \left( p\!\!\!/ + m \right) | q + \r,
& & 
v(p,-) = \frac{1}{\sqrt{2 p q}} \left( p\!\!\!/ - m \right) | q + \r.
\end{eqnarray}
For the conjugate spinors we have
\begin{eqnarray}\label{eq:5}
\bar{u}(p,+) = \frac{1}{\sqrt{2 p q}} \l q - | \left( p\!\!\!/ + m \right), 
& & 
\bar{v}(p,+) = \frac{1}{\sqrt{2 p q}} \l q - | \left( p\!\!\!/ - m \right),
\nonumber \\
\bar{u}(p,-) = \frac{1}{\sqrt{2 p q}} \l q + | \left( p\!\!\!/ + m \right),
& & 
\bar{v}(p,-) = \frac{1}{\sqrt{2 p q}} \l q + | \left( p\!\!\!/ - m \right). 
\end{eqnarray}
It is easy to check that for these spinors the Dirac equations,
orthogonality, and completeness relations hold.  The dependency on the
arbitrary reference momentum $q$ drops out in the final answer.

Given two four-vectors $p$ and $q$, the spinor product
$\l p q \r$ is calculated as follows:
If $p_t > 0$ and $q_t > 0$,
\bq
     \langle pq \rangle = \frac{1}{\sqrt{p_+ q_+}} \left( p_\perp q_+ - p_+ q_
\perp \right)\,.
\eq
Here the light-cone coordinates $p_+ = p_t + p_z$ and $p_\perp = p_x +
i p_y$ are used.  For negative-energy four-vectors we have
\bq
     \langle p q \rangle & = & - \langle (-p) q \rangle, \;\;\; \mbox{for} \;
p_t < 0\,, \nonumber \\ 
     \langle p q \rangle & = & - \langle p (-q) \rangle, \;\;\; \mbox{for} \;
q_t < 0\,.
\eq
The spinor product $[ p q ]$ is related to $\l p q \r$ by
\bq
     [ p q ] = \mbox{sign}(p_t q_t) \langle qp \rangle^\ast\,.
\eq

We employ amplitudes with all partons outgoing, generating in an
economical way the relevant scattering amplitudes by crossing.
The lowest-order amplitude is shown in Fig.~\ref{fig:1}
\begin{eqnarray}
\label{eq:2}
A_{Wb} & : & 0 \rightarrow t(p_8) + \bar{b}(p_4) + d(p_6) + \bar{u}(p_7)\,,
\end{eqnarray}
where all momenta are outgoing.
We use here the notation of Ref.~\cite{vanderHeide:2000fx},
which explains the unusual labeling of the 
momenta with $p_4$, $p_6$, $p_7$ and $p_8$.

Each amplitude we decompose into gauge-invariant partial amplitudes.
The color decomposition of the Born amplitude is
\begin{eqnarray}
A_{Wb,born} & = & \delta_{84} \delta_{67} A_{Wb,born}^{[1]}\,.
\end{eqnarray}
It is convenient to factor out some common prefactors 
from the partial
amplitude $A_{Wb,born}^{[1]}$ and to write it as follows:
\begin{eqnarray}
A_{Wb,born}^{[1]} & = & \frac{e^2 V_{ud}^\ast V_{tb}}{2 \sin^2 \theta_W} \cdot
\frac{2 i}{s_{67} -m_W^2}
\frac{B^{[1]}_{Wb,born}}{\sqrt{ - \l 2- | 4+6+7 |2- \r}}.
\end{eqnarray}
Here we denote the reference momentum for the massive spinor by $q=p_2$.
If one is only interested in the spin-summed squared amplitude, one may
choose any arbitrary null vector for $p_2$, the choices $p_2=(1,0,0,1)$ 
or $p_2=p_6$ are examples.
However, by keeping $p_2$ unspecified, we keep the complete spin information,
and our formulas become only slightly more lengthy.
The non-vanishing Born amplitudes are
\begin{eqnarray}
B^{[1]}_{Wb,born}(p_4^+,p_6^-,p_7^+,p_8^-) & = & [ 47 ] \l 6- | 4+7 | 2- \r, \nonumber \\
B^{[1]}_{Wb,born}(p_4^+,p_6^-,p_7^+,p_8^+) & = & m \l 26 \r [ 74 ].
\end{eqnarray}

We now turn our attention to the loop amplitudes.
The color decomposition of the one-loop amplitude is given by
\begin{eqnarray}
A_{Wb,loop} & = & \delta_{67} T^a_{8i} T^a_{i4} A_{Wb,loop}^{[1]}
 + T^a_{6i} T^a_{i7} \delta_{84} A_{Wb,loop}^{[2]}
 + T^a_{67} T^a_{84} A_{Wb,loop}^{[3]} \nonumber \\
& = & \frac{N^2-1}{2N} \delta_{67} \delta_{84} A_{Wb,loop}^{[1]}
 + \frac{N^2-1}{2N} \delta_{67} \delta_{84} A_{Wb,loop}^{[2]} \nonumber \\
& &  + \frac{1}{2} \left( \delta_{64} \delta_{87}
- \frac{1}{N} \delta_{67} \delta_{84} \right) A_{Wb,loop}^{[3]}.
\end{eqnarray}
Here we used the short-hand notation
\begin{eqnarray}
T^a_{84} & = & T^{a}_{i_8 j_4}\,,
\end{eqnarray}
where $a$ is the color index of the gluon, $i_8$ is the color
index of the quark $t(p_8)$ and $j_4$ is the color index of the quark
$\bar{b}(p_4)$.
$A_{Wb,loop}^{[1]}$ corresponds to loop corrections on the $t$-$b$ line,
$A_{Wb,loop}^{[2]}$ to corrections on the $u$-$d$ line, and
$A_{Wb,loop}^{[3]}$ to a gluon exchange between the two lines.
Note that we do not have to calculate $A_{Wb,loop}^{[3]}$:
\begin{eqnarray}
2 \; \mbox{Re}\; \left( \delta_{84} \delta_{67} A_{Wb}^{[1]} \right)^\ast 
T^a_{67} T^a_{84} A_{Wb,loop}^{[3]} & = & 0 \,,
\end{eqnarray}
because the color matrices are traceless.  We write
\begin{eqnarray}
A^{[1]}_{Wb,loop} & = & \frac{e^2 V_{ud}^\ast V_{tb}}{2 \sin^2 \theta_W} 
\frac{2i}{s_{67} -m_W^2} \frac{B^{[1]}_{Wb,loop}}{\sqrt{-\l 2- |4+6+7|2- \r }} \frac{g^2}{(4 \pi)^2} \nonumber \\
A^{[2]}_{Wb,loop} & = & \frac{e^2 V_{ud}^\ast V_{tb}}{2 \sin^2 \theta_W} 
\frac{2i}{s_{67} -m_W^2} \frac{B^{[2]}_{Wb,loop}}{\sqrt{-\l 2- |4+6+7|2- \r }} \frac{g^2}{(4 \pi)^2} \,.
\end{eqnarray}

For the helicity configuration $p_4^+, p_6^-, p_7^+, p_8^-$ we obtain
\begin{eqnarray}
\lefteqn{
B^{[1]}_{Wb,loop}(p_4^+,p_6^-,p_7^+,p_8^-) = } & & \nonumber \\
& & 2 [ 47 ] \Biggl\{ \l 6- | 4+7 | 2- \r \l 4- | 6+7 | 4- \r 
C_0^{(a)}(s_{67},m^2) \nonumber \\
& & + \Biggl[ \Biggl( - \frac{1}{2} + \frac{s_{67}}{\l 4- | 6+7 | 4- \r }
\Biggr) \l 6- | 4+7 | 2- \r + \frac{1}{2} \frac{s_{467}}{\l 4- | 6+7 | 4- \r }
\l 64 \r [42] \Biggr] B_0^{(a)}(s_{67},m^2) \nonumber \\
& & - \frac{1}{\l 4- | 6+7 | 4- \r} ( s_{67} \l 6- | 4+7 | 2- \r + s_{467} \l
64 \r [42] ) B_0^{(b)}(m^2) \nonumber \\
& & + \frac{1}{2} \frac{\l 64 \r [42]}{\l 4- | 6+7 | 4- \r} A_0(m^2)
- s_{467} \frac{\l 64 \r [42]}{\l 4- | 6+7 | 4- \r} C_0^{(-2\eps)}(s_{67},m^2)
\Biggr\} \nonumber \\
\lefteqn{
B^{[2]}_{Wb,loop}(p_4^+,p_6^-,p_7^+,p_8^-) = } \nonumber \\
& & -2 [ 47 ] \l 6- | 4+7 | 2- \r \biggl( s_{67} C_0^{(b)}(s_{67}) 
+ \frac{3}{2} B_0^{(c)}(s_{67}) \biggr) \,.
\end{eqnarray}

For the helicity configuration $p_4^+, p_6^-, p_7^+, p_8^+$ we obtain
\begin{eqnarray}
\lefteqn{
B^{[1]}_{Wb,loop}(p_4^+,p_6^-,p_7^+,p_8^+) = } & & \nonumber \\
& & -2m [ 74 ] \Biggl[ \l 62 \r \l 4- | 6+7 | 4- \r C_0^{(a)}(s_{67},m^2) 
\nonumber \\
& & + \frac{1}{\l 4- | 6+7 | 4- \r} \Biggl( s_{67} \l 62 \r 
 - \frac{1}{2} \l 67 \r [ 74 ] \l 42 \r \Biggr)
 B_0^{(a)}(s_{67},m^2) \nonumber \\
& & - \frac{1}{\l 4- | 6+7 | 4- \r} ( s_{67} \l 62 \r + \l 64 \r \l 4+
| 6+7 | 2+ \r )
 B_0^{(b)}(m^2) \nonumber \\
& & + \frac{1}{2} \frac{\l 64 \r \l 4+ | 6+7 | 2+ \r}{s_{467} \l 4- | 6+7 | 4-
\r} A_0(m^2)
+ \frac{ \l 2- | 6+7 | 4- \r \l 46 \r}{\l 4- | 6+7 | 4- \r}
C_0^{(-2\eps)}(s_{67},m^2)
\Biggr] \nonumber \\
\lefteqn{
B^{[2]}_{Wb,loop}(p_4^+,p_6^-,p_7^+,p_8^+) = } \nonumber \\
& & -2m [ 74 ] \l 26 \r \biggl( s_{67} C_0^{(b)}(s_{67}) + \frac{3}{2}
B_0^{(c)}(s_{67}) \biggr) \,.
\end{eqnarray}
The expressions for the standard scalar integrals are collected in the
Appendix.  The ultraviolet (UV) renormalization is discussed in the
next section.

Finally, we need the real emission amplitudes with one additional
gluon.  These are listed in Eqs.~(\ref{eq:real1}--\ref{eq:real4}), and
correspond to the process
\begin{eqnarray}
\label{reaction2}
A_{Wg} & : & 0 \rightarrow t(p_8) + \bar{b}(p_4) + g(p_5) + d(p_6) +
\bar{u}(p_7) \,.
\end{eqnarray}
The color decomposition reads
\begin{eqnarray}
A_{Wg,real} & = & T^5_{84} \delta_{67} A_{Wg,real}^{[1]}
    + \delta_{84} T^5_{67} A_{Wg,real}^{[2]} \,.
\end{eqnarray}
These amplitudes have been calculated in Ref.\ \cite{vanderHeide:2000fx}.
For convenience we list them here.  We have
\begin{eqnarray}
A^{[1]}_{Wg,real} & = & \frac{g e^2 V_{ud}^\ast V_{tb}}{2 \sin^2 \theta_W} \cdot
\frac{(-i) 2 \sqrt{2}}{s_{67} -m_W^2}
\frac{B^{[1]}_{Wg,real}}{\sqrt{ - \l 2- | 4+5+6+7 |2- \r}}, \nonumber \\
A^{[2]}_{Wg,real} & = & \frac{g e^2 V_{ud}^\ast V_{tb}}{2 \sin^2 \theta_W} \cdot
\frac{(-i) 2 \sqrt{2}}{s_{567} -m_W^2}
\frac{B^{[2]}_{Wg,real}}{\sqrt{ - \l 2- | 4+5+6+7 |2- \r}}.
\end{eqnarray}
As reference momentum for the massive spinor we have chosen $q=p_2$.
The non-vanishing amplitudes are
\begin{eqnarray}
\lefteqn{
B^{[1]}_{Wg,real}(p_4^+,p_5^+,p_6^-,p_7^+,p_8^-) } \nonumber \\
& = & 
\frac{\l 6- | 4+5+7 | 2- \r}{\l 65 \r} 
\left( \frac{\l 6- | 4+5 | 7- \r}{\l 45 \r} + \frac{[ 74 ] \l 6- | 4+7 | 5- \r}{s_{467} -m^2} \right),
\nonumber \\ 
\lefteqn{
B^{[1]}_{Wg,real}(p_4^+,p_5^+,p_6^-,p_7^+,p_8^+) } \nonumber \\
& = & 
- \frac{m \l 26 \r}{\l 65 \r}
\left( \frac{\l 6- | 4+5 | 7- \r}{\l 45 \r} + \frac{[ 74 ] \l 6- | 4+7 | 5- \r}{s_{467} -m^2} \right),
\nonumber \\ 
\lefteqn{
B^{[1]}_{Wg,real}(p_4^+,p_5^-,p_6^-,p_7^+,p_8^-) } \nonumber \\
& = &
\frac{[ 74 ]}{ [ 54 ] \left( s_{467} - m^2 \right)}
\left( \l 5- | 4+6+7 | 2- \r [ 47 ] \l 76 \r + m^2 [ 24 ] \l 56 \r \right),
\nonumber \\ 
\lefteqn{
B^{[1]}_{Wg,real}(p_4^+,p_5^-,p_6^-,p_7^+,p_8^+) } \nonumber \\
& = &
- \frac{m}{s_{467} -m^2} \frac{ [ 47 ] }{ [ 45 ] }
\left( \l 25 \r \l 67 \r [ 74 ] + \l 56 \r \l 2- | 5+6+7 | 4- \r \right),
\end{eqnarray}
\begin{eqnarray}
B^{[2]}_{Wg,real}(p_4^+,p_5^+,p_6^-,p_7^+,p_8^-) & = & 
\frac{\l 6- | 4+5+7 | 2- \r \l 6- | 5+7 | 4- \r}{\l 56 \r \l 75 \r}, \nonumber \\
B^{[2]}_{Wg,real}(p_4^+,p_5^+,p_6^-,p_7^+,p_8^+) & = & 
\frac{m \l 62 \r \l 6- | 5+7 | 4- \r}{\l 56 \r \l 75 \r}, \nonumber \\
B^{[2]}_{Wg,real}(p_4^+,p_5^-,p_6^-,p_7^+,p_8^-) & = & 
\frac{[ 74 ] \l 2+ | (4+5+6+7) ( 5+6) | 7- \r}{ [ 57 ] [ 56 ]}, \nonumber \\
B^{[2]}_{Wg,real}(p_4^+,p_5^-,p_6^-,p_7^+,p_8^+) & = & 
\frac{m [ 47 ] \l 2- | 5+6 | 7- \r}{ [ 57 ] [ 56 ]}.
\end{eqnarray}

The matrix element squared is given by
\begin{eqnarray}
\left| A_{Wg,Real} \right|^2 & = & \frac{1}{2} N_C ( N_C^2-1) \left( 
\left| A_{Wg,Real}^{[1]} \right|^2 + \left| A_{Wg,Real}^{[2]} \right|^2 \right)
\,.
\end{eqnarray}
There are no interference terms between $A_{Wg,Real}^{[1]}$ and
$A_{Wg,Real}^{[2]}$.

\subsection{Conversion between schemes and scheme independence}
\label{sec:scheme}

The one-loop amplitudes presented in the previous section have been
calculated in a four-dimensional scheme. They differ from the
corresponding amplitudes in the 't~Hooft-Veltman scheme by finite
terms.  These finite terms are either of ultraviolet (UV) or infrared
(IR) origin and result from expressions of the form $\eps/\eps$.  To
obtain the unique and correct result, one proceeds through the
following steps:
\begin{itemize}
\item The bare one-loop amplitudes may contain UV-divergences.  With
the help of a specific regularization scheme these divergences are
isolated and removed by renormalization. The explicit form of the
renormalization depends on the chosen renormalization scheme.
\item The specific combination of regularization and renormalization
scheme may break certain Ward identities. These Ward identities have
to be restored through finite renormalizations. The required finite
renormalizations are universal, i.e.~they do not depend on the process
under consideration.  After this step all finite parts of UV-origin
are uniquely fixed.
\item In addition, QCD amplitudes may contain IR-divergences.
Unitarity requires that we employ the same regularization scheme in
the phase space integral over the unresolved real emission part as in
the one-loop integral.\\ Alternatively, since the structure of the
IR-divergences is universal, we may derive simple formulas, which
relate the finite parts specific to a certain regularization scheme to
the ones of another scheme.  We can therefore convert a one-loop
amplitude calculated in one scheme to the corresponding amplitude
calculated in another scheme.
\end{itemize}

We will discuss the three steps in detail for the case at hand.  To
start, let us briefly summarize the properties of the 't~Hooft-Veltman
scheme and the four-dimensional scheme.  The 't~Hooft-Veltman scheme
treats unobserved particles (particles in loops and unresolved partons
in the real emission part) in $D=4-2\eps$ dimensions.  Observed
particles are taken in four dimensions.  $\gamma_5$ is a
four-dimensional object in the t' Hooft-Veltman scheme, anti-commuting
with the first four Dirac matrices and commuting with the remaining
ones.

The four-dimensional scheme is specified in simple terms by the
fermion propagators
\begin{eqnarray}
i \frac{p\!\!\!/_{(4)}+m 1_{(4)}}{p^2_{(D)}-m^2}\,.
\end{eqnarray}
Four-dimensional Dirac-matrices occur in the numerator, whereas $D$-dimensional
quantities occur in the denominator.
Two adjacent Dirac-matrices in the numerator are contracted as
\begin{eqnarray}
p\!\!\!/_{(4)} p\!\!\!/_{(4)} & = & \left( p^2_{(D)} - p^2_{(-2\eps)} \right)
\cdot 1_{(4)} \,,
\end{eqnarray}
which can be interpreted as the statement ``$D$ is effectively larger
then 4''.  $p_{(D)}^2$ can cancel a propagator, whereas
$p_{(-2\eps)}^2$ will give rise to an integral in $6-2\eps$
dimensions.  It should be kept in mind that the specification given
here is just a simple prescription relevant to practical
calculations. The scheme is rigorously defined in Ref.\
\cite{Weinzierl:1999xb}.

$A_{Wb,loop}$ contains ultraviolet and infrared divergences.
Ultraviolet divergences are removed after renormalization of the quark fields
\begin{eqnarray}
\psi_{bare} = Z_{\psi}^{1/2} \psi_{ren} \,.
\end{eqnarray}
The renormalized amplitude is obtained as
\begin{eqnarray}
A_{Wb,loop,ren} & = & \left( Z_\psi^{1/2} \right)^4 A_{Wb,loop,bare} \,.
\end{eqnarray}

We have to renormalize the fields such that the residuum of the
propagators is $1$. For light quarks the appropriate renormalization
constant is $1$, due to a cancellation of UV- and IR- divergent parts.
In more detail we have
\begin{eqnarray}
Z_{\psi,onshell,FD}^{1/2} & = & 1 + \frac{1}{2} \frac{g^2}{(4\pi)^2} C_F 
\left( -\Delta_{UV} + \Delta_{IR} \right), \nonumber \\
Z_{\psi,onshell,HV}^{1/2} & = & 1 + \frac{1}{2} \frac{g^2}{(4\pi)^2} C_F 
\left( -\Delta_{UV} +1_{UV} + \Delta_{IR} - 1_{IR} \right),
\end{eqnarray}
where $\Delta = 1/\eps-\gamma+\ln 4\pi$.  Here we have indicated with
a subscript UV or IR the origins of the divergent parts as well as the
origin of additional finite terms which arise from a cancellation of
$1/\eps$-terms with terms of order $\eps$.  For massive quarks we have
\begin{eqnarray}
Z_{\psi,onshell,FD}^{1/2} & = & 1 + \frac{1}{2} \frac{g^2}{(4\pi)^2} C_F 
\left( -3 \Delta_{UV} -5 + 3 \ln \frac{m^2}{\mu^2} \right) \nonumber \\
Z_{\psi,onshell,HV}^{1/2} & = & 1 + \frac{1}{2} \frac{g^2}{(4\pi)^2} C_F 
\left( -3 \Delta_{UV} +1_{UV} -5 + 3 \ln \frac{m^2}{\mu^2} \right) \,.
\end{eqnarray}

After renormalization of the quark field we have
\begin{eqnarray}
B^{[1]}_{Wb,loop,ren,FD} & = & B^{[1]}_{Wb,loop,bare,FD} + \frac{1}{2} \left(
- 3 \Delta - 5 + 3 \ln \frac{m^2}{\mu^2} \right)
B^{[1]}_{Wb,born
} \nonumber \\
B^{[2]}_{Wb,loop,ren,FD} & = & B^{[2]}_{Wb,loop,bare,FD} \,.
\end{eqnarray}
The only divergences left in the renormalized amplitudes are of
infrared origin.  The infrared divergent parts are given by
\begin{eqnarray}
\left. B^{[1]}_{Wb,loop} \right|_{IR} & = & 
-2 \Delta\left( \frac{1}{2} \Delta + \frac{1}{2} \gamma - \frac{1}{2} \ln 4
 \pi + \frac{5}{4} 
 - \ln \frac{m^2-s_{67}}{\mu^2} + \frac{1}{2} \ln \frac{m^2}{\mu^2} 
\right)  B_{Wb,born} \nonumber \\
\left. B^{[2]}_{Wb,loop} \right|_{IR} & = & 
-2 \Delta \left( \Delta + \gamma - \ln 4 \pi + \frac{3}{2} - \ln
  \frac{-s_{67}}{\mu^2} \right)  B_{Wb,born} \,.
\end{eqnarray}

We now turn our attention to Ward identities and finite
renormalizations.  In the 't~Hooft-Veltman scheme $\gamma_5$
anti-commutes with the first four Dirac matrices and commutes with
the remaining ones.  This treatment violates a Ward identity, which
has to be restored by a finite renormalization.  Explicitly, one
splits the left-handed interaction into a vector ($\gamma_\mu$) and
axial-vector [$\Gamma_{\mu 5} = 1/2 ( \gamma_\mu \gamma_5 - \gamma_5
\gamma_\mu)$] part.
\begin{eqnarray}
\gamma_\mu \frac{1}{2} \left( 1 -\gamma_5 \right) & = & \frac{1}{2} \gamma_\mu
 - \frac{1}{4} \left( \gamma_\mu \gamma_5
- \gamma_5 \gamma_\mu \right) \,.
\end{eqnarray}
The Ward identity is restored by a finite renormalization of the
axial-vector coupling:
\begin{eqnarray}
\Gamma_{\mu 5}^{bare} & = & Z_{axial,HV,fin} \Gamma_{\mu 5}^{renorm}
\nonumber \\
Z_{axial,HV,fin} & = & 1 + 4 \frac{g^2}{(4 \pi)^2} C_F \,.
\end{eqnarray}

There is some freedom in how we continue the left-handed coupling in the
't~Hooft-Veltman scheme to $D$ dimensions.  For example, the expressions
\begin{eqnarray}
\gamma_\mu \frac{1}{2} \left( 1 -\gamma_5 \right), \;\; \frac{1}{2} \left( 1 +
\gamma_5 \right) \gamma_\mu, \;\;
\frac{1}{2} \left( 1 +\gamma_5 \right) \gamma_\mu \frac{1}{2} \left( 1 -
\gamma_5 \right), \;\;
\frac{1}{2} \gamma_\mu - \frac{1}{4} \left( \gamma_\mu \gamma_5
- \gamma_5 \gamma_\mu \right) \,,
\end{eqnarray}
all agree in four dimensions, but differ in $D$ dimensions.  Of course
this difference is compensated by the appropriate finite
renormalization.  $Z_{axial,HV,fin}$ corresponds to the choice
$-1/4(\gamma_\mu \gamma_5 - \gamma_5 \gamma_\mu)$.

The four-dimensional scheme violates a Ward identity as well, which is
restored by a finite renormalization of the left-handed coupling
$\Gamma_{left} = 1/2 \gamma_\mu ( 1 - \gamma_5)$,
\begin{eqnarray}
\Gamma_{left}^{bare} & = & Z_{EW,FDfin} \Gamma_{left}^{renorm} \nonumber \\
Z_{EW,FD,fin} & = & 1 + \frac{g^2}{(4 \pi)^2} C_F \,.
\end{eqnarray}
After the finite renormalization the renormalized amplitudes
\begin{eqnarray}
A_{Wb,loop,ren,HV} & = & \left( Z_{\psi,HV}^{1/2} \right)^4 A_{Wb,loop,HV}
\nonumber \\
A_{Wb,loop,ren,FD} & = & \left( Z_{\psi,FD}^{1/2} \right)^4  
\left( Z_{EW,FD,fin} \right)^{-2} A_{Wb,loop,FD} \,,
\end{eqnarray}
agree up to terms resulting from a different treatment of the infrared
divergences.  The axial-vector coupling in the 't~Hooft-Veltman scheme
has been renormalized according to
\begin{eqnarray}
\Gamma_{\mu 5}^{bare} & = & Z_{axial,HV,fin} \Gamma_{\mu 5}^{renorm} \,.
\end{eqnarray}

In detail, we have for the finite parts of UV-origin in terms of
$g^2/(4\pi)^2 C_F B^{Born}$
\begin{eqnarray}
\begin{array}{|c|c|c|c|}
\hline
 & FD & \mbox{HV}_{vector} & \mbox{HV}_{axial} \\
B_{Wb,loop,bare} & 0 & -2 & +2 \\
Z_{coupling}^{-1} & -1 & 0 & -4 \\
Z_\psi & 0 & +1 & +1 \\
\hline
\mbox{Sum} & -1 & -1 & -1 \\
\hline
\end{array}
\end{eqnarray}

The remaining differences are due to finite terms of infrared origin.
Due to the universal structure of the infrared divergences we may
relate the amplitudes calculated in the four-dimensional scheme to the
ones in the 't~Hooft-Veltman scheme
\cite{Kunszt:1994sd,Catani:1997pk,Phaf:2001gc}.  The relations are
\begin{eqnarray}
B_{Wb,loop,HV}^{[1]} & = & B_{Wb,loop,FD}^{[1]} - \frac{1}{2} B_{Wb,Born}^{[1]}
\nonumber \\ 
B_{Wb,loop,HV}^{[2]} & = & B_{Wb,loop,FD}^{[2]} - B_{Wb,Born}^{[1]} \,.
\end{eqnarray}

This completes our discussion on scheme-independence.  In summary, we
are able to perform the calculation in a four-dimensional scheme and
to obtain from this result the amplitudes in the 't~Hooft-Veltman
scheme through simple and universal relations.

We would like to comment on the original formulation of dimensional
reduction.  Dimensional reduction \cite{Siegel:1979wq,Siegel:1980qs}
differs from the four-dimensional scheme by use of the relation
\begin{eqnarray}
p\!\!\!/_{(4)} p\!\!\!/_{(4)} & = & p^2_{(D)} \cdot 1_{(4)} \,,
\end{eqnarray}
which can be interpreted as ``D is smaller than 4''.  The results in
dimensional reduction can easily obtained from ours by dropping the
terms $C_0^{(-2\eps)}$.  The $C_0^{(-2\eps)}$-terms yield finite terms
related to UV-divergences.  We note that the results in dimensional
reduction cannot be related to ours, nor to the 't~Hooft-Veltman
scheme by a finite renormalization, since the term with
$C_0^{(-2\eps)}$ in $B_{Wb,loop}^{[1]}$ is not proportional to
$B_{Wb,born}$.  Therefore the naive approach of dimensional reduction
is not consistent.  The situation may be cured at the expense of
introducing additional scalar ghost particles \cite{Capper:1980ns},
however this spoils the calculational simplicity of the scheme.  From
a calculational point of view we prefer the scheme defined in Ref.\
\cite{Weinzierl:1999xb}, since $6-2\eps$-dimensional integrals are
rather ``inexpensive'' to evaluate.

\subsection{Spin observables}

The helicity amplitudes contain the complete spin information for the
single-top-quark processes.  In the old-fashioned method, spin
observables are calculated by inserting the spin projection operator
\bq
\label{spinprojector}
u(p,s) \bar{u}(p,s) & = & \left( p\!\!\!/ + m \right) 
 \frac{1}{2} \left( 1 + \gamma_5 s\!\!\!/ \right) \,,
\eq
into the matrix element squared. In Eq.~(\ref{spinprojector}) $s$
denotes a spin four-vector with $s^2 = -1$, and $ p \cdot s = 0$.  In the
rest frame of the particle the spatial components of $s$ point in the
same direction as the spin of the particle \cite{Mahlon:1996zn}.  To
make contact with this formulation, we first introduce the spin
density matrix \cite{Brandenburg:1998xw} in the basis of
Eq.~(\ref{eq:4}):
\bq
\rho & = & \left( 
\begin{array}{cc}
A(...,p^+,...) A^\ast(...,p^+,...) & A(...,p^+,...) A^\ast(...,p^-,...) \\ 
A(...,p^-,...) A^\ast(...,p^+,...) & A(...,p^-,...) A^\ast(...,p^-,...) \\ 
\end{array}
\right) \,.
\eq
In addition we need the projection operator Eq.~(\ref{spinprojector})
in the basis of Eq.~(\ref{eq:4}). This one is obtained as
\bq
P & = & \frac{1}{2} \frac{1}{2 p q} \left( 
\begin{array}{cc}
-\l q- | (p\!\!\!/ + m ) (1 - s\!\!\!/ ) | q- \r &
 \l q- | (p\!\!\!/ + m ) (1 - s\!\!\!/ ) | q+ \r \\
-\l q+ | (p\!\!\!/ + m ) (1 - s\!\!\!/ ) | q- \r &
 \l q+ | (p\!\!\!/ + m ) (1 - s\!\!\!/ ) | q+ \r \\
\end{array}
\right) \,.
\eq

Spin observables are then calculated as
\bq
\label{spinobservable}
\mbox{Tr}\; \left( P \cdot \rho \right).
\eq
Note that the entries of the matrices are complex numbers, and that the
spin vector $s$ enters only through the matrix $P$.  It is easily
verified that this expression agrees with the one obtained from
Eq.~(\ref{spinprojector}).  The spin summed result is recovered by
replacing $P$ in Eq.~(\ref{spinobservable}) with the unit matrix.

\section{Numerical results}
\label{sec:numerical}

The inclusive NLO cross sections for $s$- and $t$-channel production
of single-top-quarks were published in Refs.~\cite{Smith:1996ij} and
\cite{Stelzer:1997ns}, respectively.  After we choose our numerical
inputs, we compare to these older calculations, and update the
inclusive cross sections using newer parton distribution functions
(PDFs).

In order to make a definite comparison to the older calculations, we
reevaluate all results with the following parameters: For the mass of
the $W$ boson we use $m_W=80.4$ GeV.  For the top-quark mass we take
$m_t=175$ GeV.  In LO calculations we use CTEQ5L PDFs
\cite{Lai:1999wy}.  In NLO calculations we use CTEQ5M1 PDFs with
2-loop running of $\alpha_s$.  We define the electroweak coupling by
$g^2 = 8 G_F M_W^2 / \sqrt{2}$, with Fermi coupling constant of $G_F =
1.16639 \times 10^{-5}$ GeV$^{-2}$.  We consider $p\bar{p}$ collisions
with center-of-momentum energy $\sqrt{S}=1.8$, $1.96$, or $2.0$ TeV
(Tevatron), and $p p$ collisions with a center-of-momentum energy
$\sqrt{S}=14$ TeV (LHC).

\begin{figure}[htbp]
\centerline{
\epsfig{file=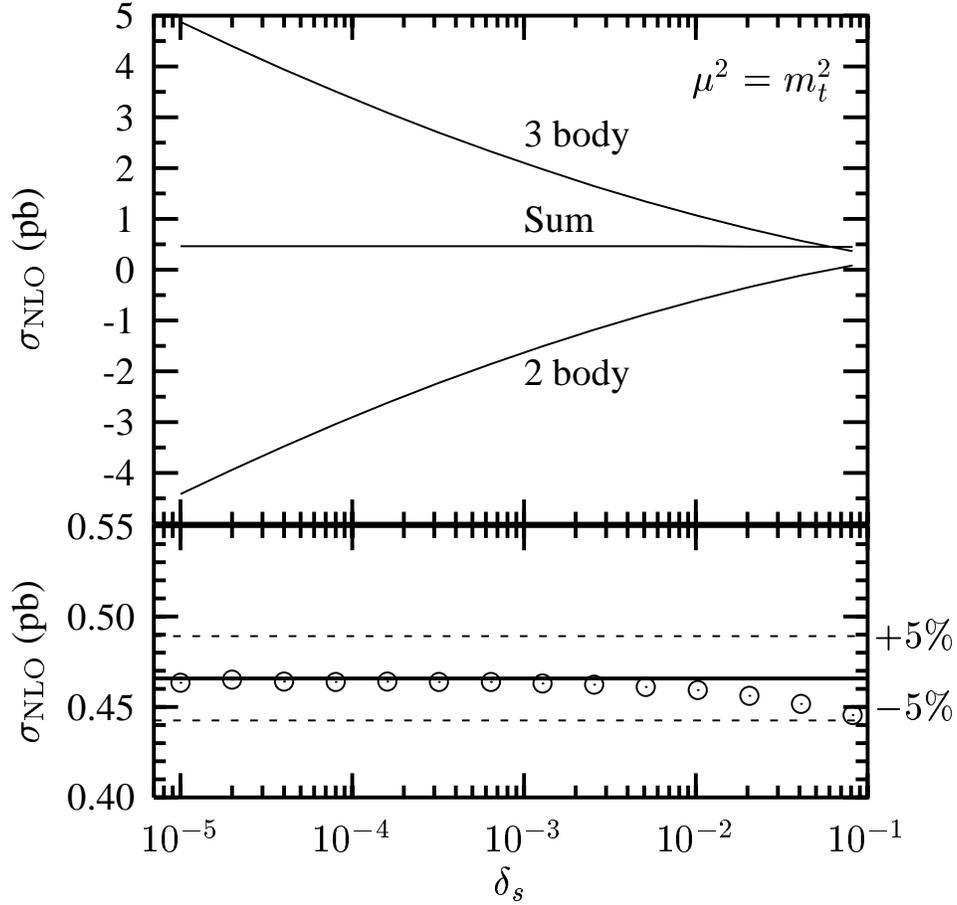, width=5in} } 
\caption{\small The next-to-leading order 
single-top-quark total cross section for the $s$-channel 
at a $\sqrt{S}=2$ TeV proton-antiproton machine.
The two- and three- body contributions, together with their sum, 
are shown as a function of the soft cutoff $\delta_s$.  The 
bottom enlargement shows the sum (open circles) relative to 
$\pm 5 \%$ (dotted lines) of the analytic result (solid line).
\label{fig:pss2s}}
\end{figure}

\begin{figure}[htbp]
\centerline{
\epsfig{file=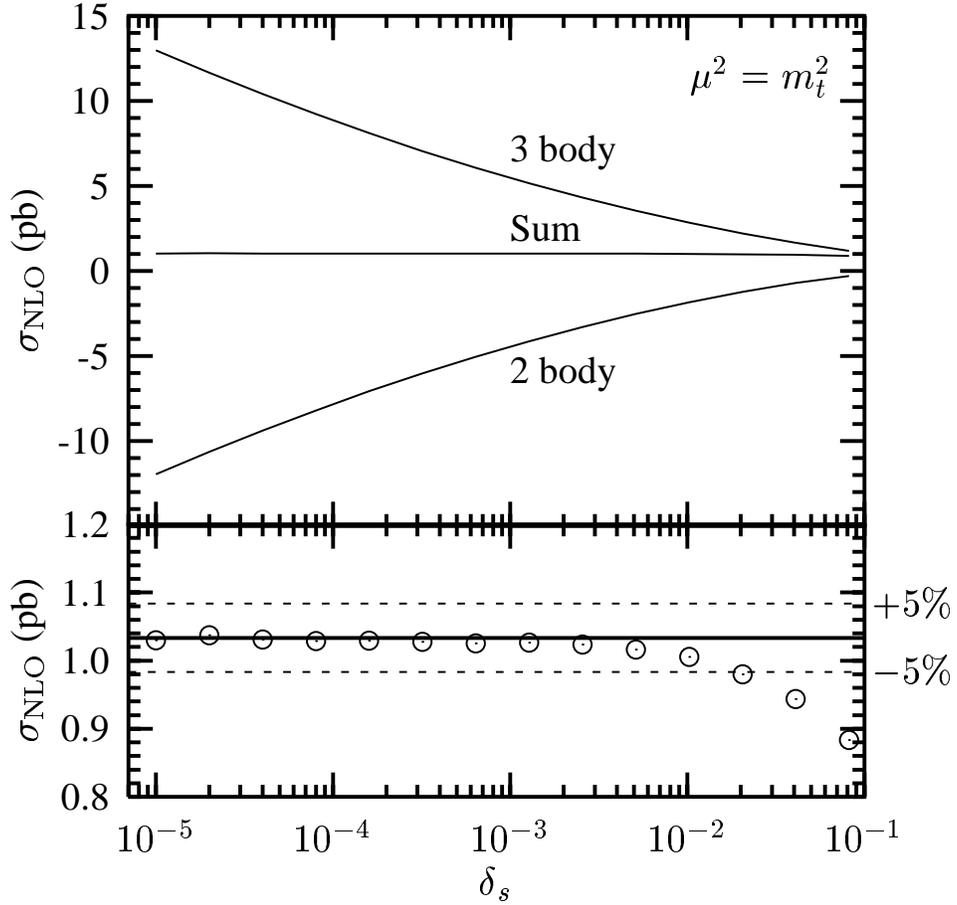, width=5in} } 
\caption{\small The next-to-leading order single-top-quark total cross
section for the $t$-channel at a $\sqrt{S}=2$ TeV proton-antiproton
machine.  The two- and three- body contributions, together with their
sum, are shown as a function of the soft cutoff $\delta_s$.  The
bottom enlargement shows the sum (open circles) relative to $\pm 5\%$
(dotted lines) of the analytic result (solid line).\label{fig:pss2t}}
\end{figure}

Unlike the massive dipole formalism, the phase space slicing method
of two cutoffs depends on explicit parameters $\delta_s$ and
$\delta_c$.  In Figs.~\ref{fig:pss2s} and \ref{fig:pss2t} we show the
$s$- and $t$-channel inclusive cross sections as a function of
$\delta_s = 300\times\delta_c$.  The logarithmic dependence cancels in
the sum of the two- and three-body contributions and leaves terms
proportional to $\delta_s$ and $\delta_c$.  By taking $\delta_s < $ a
few $\times 10^{-3}$, the cross sections converge to the updated
analytic results.

The LO and NLO cross sections for $s$-channel and $t$-channel
single-top-quark production are listed in picobarns in Tables
\ref{tab:sigmt} and \ref{tab:sigtot}.  Since the Tevatron is a $p\bar
p$ collider, the cross sections for top-quark ($t$) and antitop-quark
($\bar t$) production are the same.  The LHC is a $pp$ collider, and
hence the $t$ and $\bar t$ cross sections are listed separately.
Factorization ($\mu_F$) and renormalization ($\mu_R$) scales are set
equal.  In Table \ref{tab:sigmt} all scales are set to the top mass
$m_t$.  In Table \ref{tab:sigtot} the $s$-channel cross sections are
calculated using the invariant mass of the top-quark/antibottom-jet
pair for the scale, $\mu^2 = M^2_{t\bar b} = (P_t+P_{\bar b})^2$,
where $P_t$ and $P_{\bar b}$ are the four-momenta of the top quark and
antibottom jet, respectively.\footnote{ The choice of jet definition
induces a cone-size dependence in the scale that always enters the
cross section at one higher order in QCD.  We have confirmed that this
effect is numerically less than the overall scale uncertainty for any
infrared-safe jet definition.}  The $t$-channel cross section uses the
double deep-inelastic-scattering (DDIS) scales, $\mu_l^2 = Q^2 =
-(p_b-p_t)^2$ for the light-quark line and $\mu_h^2 = Q^2 + m_t^2 =
-(p_u-p_d)^2 + m_t^2$ for the heavy-quark line, where $Q^2$ is the
virtuality of the exchanged $W$ boson (valid through NLO), and $p_i$
are the four-momenta of the partons in Eq.~(\ref{eq:3}).  All LO cross
sections are identical to those in Refs.~\cite{Smith:1996ij} and
\cite{Stelzer:1997ns} once updated to the CTEQ5 PDFs.

\begin{table}[tb]
\caption{\small LO and NLO cross sections for single-top-quark
production at the Tevatron and LHC for $m_t=175$ GeV.  Cross sections
are evaluated with CTEQ5L and CTEQ5M1 PDFs, and all scales set to
$m_t$.  Errors include only Monte Carlo statistics.\label{tab:sigmt}}
\begin{center}
\begin{tabular}{c|r@{~}l|c|r@{~}l} \hline 
Process&\multicolumn{2}{c|}{$\sqrt{S}$}& $\sigma_{LO}$ (pb) & \multicolumn{2}{c}{$\sigma_{NLO}$ (pb)} \\
\hline
$s$-channel&1.8& TeV $p\bar p$ ($t$) & 0.259 & 0.380 & $\pm$0.002 \\
&1.96& TeV $p\bar p$ ($t$) & 0.304 & 0.447 & $\pm$0.002 \\
&2& TeV $p\bar p$ ($t$) & 0.315 & 0.463 & $\pm$0.002 \\
&14& TeV $pp$ ($t$) & 4.53 & 6.55 & $\pm$0.03 \\
&14& TeV $pp$ ($\bar t $) & 2.74 & 4.07 & $\pm$0.02 \\
\hline
$t$-channel&1.8& TeV $p\bar p$ ($t$) & 0.648 & 0.702 & $\pm$0.003 \\
&1.96& TeV $p\bar p$ ($t$) & 0.883 & 0.959 & $\pm$0.002 \\
&2& TeV $p\bar p$ ($t$) & 0.948 & 1.029 & $\pm$0.004 \\
&14& TeV $pp$ ($t$) & 144.8 & 152.6 & $\pm$0.6 \\
&14& TeV $pp$ ($\bar t $) & 83.4 & 90.0 & $\pm$0.5 \\
\hline
\end{tabular}
\end{center}
\end{table}

\begin{table}[tb]
\caption{\small LO and NLO cross sections for single-top-quark
production at the Tevatron and LHC for $m_t=175$ GeV.  Cross sections
are evaluated with CTEQ5L and CTEQ5M1 PDFs, and $M_{t\bar b}$ or the
DDIS scales ($\mu_l=Q^2$, $\mu_h=Q^2 + m_t^2$), for $s$-channel or
$t$-channel, respectively.  Errors include only Monte Carlo
statistics.\label{tab:sigtot}}
\begin{center}
\begin{tabular}{c|r@{~}l|c|r@{~}l} \hline 
Process&\multicolumn{2}{c|}{$\sqrt{S}$}& $\sigma_{LO}$ (pb) & \multicolumn{2}{c}{$\sigma_{NLO}$ (pb)} \\
\hline
$s$-channel&1.8& TeV $p\bar p$ ($t$) & 0.244 & 0.377 & $\pm$0.002 \\
&1.96& TeV $p\bar p$ ($t$) & 0.287 & 0.442 & $\pm$0.002 \\
&2& TeV $p\bar p$ ($t$) & 0.297 & 0.459 &$\pm$0.002 \\
&14& TeV $pp$ ($t$) & 4.612 & 6.56 &$\pm$0.03 \\
&14& TeV $pp$ ($\bar t $) & 2.788 & 4.09 &$\pm$0.02 \\
\hline
$t$-channel&1.8& TeV $p\bar p$ ($t$) & 0.735 & 0.725 & $\pm$0.003 \\
&1.96& TeV $p\bar p$ ($t$) & 0.996 & 0.990 & $\pm$0.002 \\
&2& TeV $p\bar p$ ($t$) & 1.068 & 1.062 & $\pm$0.004 \\
&14& TeV $pp$ ($t$) & 152.7 & 155.9 & $\pm$0.6 \\
&14& TeV $pp$ ($\bar t $) & 86.1 & 90.7 & $\pm$0.5 \\
\hline
\end{tabular}
\end{center}
\end{table}

We compare the NLO $s$-channel cross sections to a recoding of
Refs.~\cite{Chang:1981qq} and \cite{Smith:1996ij}.  The results agree
to within $1\%$ for all scale choices.  Of note is our use of
$M_{t\bar b}$ for the scale in Table \ref{tab:sigtot} rather than
$Q^2$, the virtuality of the $W$, which was used in
Ref.~\cite{Smith:1996ij}.  These two scales are identical at LO and in
initial-state corrections, but differ by the emitted gluon in final
state corrections.  Since we are interested in making cuts based on
observables, such as $M_{t\bar b}$, we choose this as the scale.
While the central value of any of the scale choices is very similar,
the uncertainty is slightly larger at NLO using $m_t$ or $M_{t\bar
b}$.  In particular, if we vary the scales between $M_{t\bar b}/2$ and
$2 M_{t\bar b}$ (or $m_t/2$ and $2m_t$), we find the $s$-channel cross
section varies by $+7.8-6.9\%$ at LO, and $+5.7-5.0\%$ ($+5.5-4.6\%$)
at NLO.  In contrast, using a scale of $Q^2$, as in
Ref.~\cite{Smith:1996ij}, would predict a NLO scale uncertainty of
$\pm 4\%$.  Given that we can probe a more restricted phase space with
cuts, we take the conservative view that $\pm 5.7\%$ is an appropriate
estimate of the NLO scale uncertainty at the Tevatron when looking at
exclusive final states.  At the LHC the scale uncertainty is less than
$\pm 2\%$.

At LO and NLO, but not NNLO, color conservation forbids the exchange
of a gluon between the light- and heavy-quark lines.  Hence, the
$t$-channel process may be factorized at NLO into two independent
corrections that each resemble deep-inelastic-scattering.  The NLO
$t$-channel cross sections in Ref.~\cite{Stelzer:1997ns} were
calculated using the scales suggested by this relation to double
deep-inelastic-scattering.  When using the DDIS scales, our results in
Table \ref{tab:sigtot} match the updated Ref.~\cite{Stelzer:1997ns} to
better than $0.3\%$ at the Tevatron, but are larger by $2.9\%$ at the
LHC.  The cross sections in Table \ref{tab:sigmt} agree with the
updated results of Ref.~\cite{Stelzer:1997ns} to within $1\%$ when
evaluated at $\mu=m_t$.  In all cases, the dipole subtraction
calculations and phase space slicing calculations agree within the
statistical errors.

A subtle issue arises in attempting to ascertain the effect of higher
orders by varying the scale.  In Ref.~\cite{Stelzer:1997ns} only the
scales in the vertex and PDFs of the heavy-quark line ($b \to t$) were
varied because the corrections to the light-quark structure function
are small.\footnote{The NLO correction to the light-quark line
effectively undoes the extraction of the NLO PDFs from the DIS data.}
However, to the extent this process looks like double
deep-inelastic-scattering, we expect a similarly small effect for the
heavy-quark corrections as well.  Indeed, the results for LO and NLO
are nearly identical when using the DDIS scales, but differ
significantly when using a fixed scale such as $m_t$.  In Table
\ref{tab:tscale} we show the effects of varying the scales at the
Tevatron ($\sqrt{S}=2$ TeV) together, and separately in the light- and
heavy-quark lines.  What we see is that varying the scales together at
LO, particularly when using $m_t$, severely underestimates the NLO
correction.

\begin{table}[tb]
\caption{\small Scale variation of the LO and NLO cross sections for
$t$-channel single-top-quark production at the Tevatron ($\sqrt{S}=2$
TeV).  Variation in the light-quark and heavy-quark lines are listed
as $\mu_l$ and $\mu_h$, respectively.  Fixed ($\mu=m_t$), and double
deep-inelastic-scattering scales are shown
separately.\label{tab:tscale}}
\begin{center}
\begin{tabular}{l|c|lll} \hline 
&$\sigma_t$ & \multicolumn{1}{c}{$\mu_h$ \& $\mu_l$} &
\multicolumn{1}{c}{$\mu_h$} &\multicolumn{1}{c}{$\mu_l$} \\ \hline
\rowspace LO$_t$ ($m_t$)& $0.95$ pb & $\pm 1\%$ & $^{-7.5}_{+5.5}\%$ &
$^{+6.7}_{-5.8}\%$ \\
NLO$_t$ ($m_t$)& $1.03$ pb& $\pm 2.5\%$ & $^{-3.5}_{+4.0}\%$ & $\pm 1\%$ \\
\hline \rowspace
LO$_t$ (DDIS)& $1.07$ pb & $^{+0.1}_{-2}\%$ & $^{-7.2}_{+5.2}\%$ 
& $^{+8}_{-6.8}\%$ \\
NLO$_t$ (DDIS)& $1.06$ pb & $\pm 3.5\%$ & $^{-3}_{+4}\%$ & $\pm 0.6\%$ \\ 
\hline
\end{tabular}
\end{center}
\end{table}

The reason for the underestimate in varying the scales together is a
series of accidental cancellations that are driven by the range of
proton momentum fraction probed at the Tevatron.  For a top mass of
175~GeV and a machine of around 2~TeV, the typical $x\sim 0.1$.  For
scales also around 100--200~GeV, the $b$ PDF happens to increase with
increasing scale, whereas the valence quarks decrease.  This may be
seen in the opposite signs in the last two columns of Table
\ref{tab:tscale}.  The net effect is that to estimate the uncertainty,
we must vary the scales independently, and then add them in
quadrature.\footnote{The light- and heavy-quark corrections are
factorizable through NLO, and hence are only weakly correlated through
the evolution of the PDFs.}  This leads to a LO uncertainty of $\sim
\pm 10\%$, and a NLO uncertainty for the Tevatron of $\pm 4\%$.  At
the LHC, the NLO uncertainty is $\pm 3\%$.  These uncertainties are
consistent with expected higher-order corrections, and with both fixed
and DDIS scale choices.

While the $s$-channel cross section has only changed by a couple of
percent from \cite{Smith:1996ij}, the $t$-channel cross section is
$13\%$ smaller at the Tevatron than appears in \cite{Stelzer:1997ns}.
The shift in the NLO $t$-channel cross section is due to the
correction of bugs that appeared in the evolution of the gluon PDF in
all of the older MRS and CTEQ NLO PDFs \cite{Blumlein:1996rp}.
Because the $b$ PDF is constructed almost entirely out of gluon
evolution \cite{Stelzer:1997ns,Sullivan:2001ry}, the effect of the bug
is greatly enhanced in all processes where there is a $c$ or $b$ quark
in the initial state.  The programming bug is corrected in MRS99
(updated) \cite{Martin:1999ww} and CTEQ5M1 \cite{Lai:1999wy}.

The central goal of our calculations is not to recalculate inclusive
cross sections, but to provide full momentum and spin dependent
distributions with the option for arbitrary cuts.  Detailed analyses
of the phenomenological issues concerning these distributions will be
presented elsewhere \cite{HLPSW2}.  Here we restrict ourselves to a
comparison of the results using the phase space slicing method and
massive dipole formalism.

For our comparisons we reconstruct jets using a $k_T$ cluster
algorithm~\cite{Ellis:tq} with $\Delta R = 1$.  We define a simple
detector acceptance by assuming that only jets with $p_T > 20$~GeV and
$|\eta|<2$ are observed.  We calculate all cross sections at the scale
$\mu=m_t=175$~GeV, and at a 2~TeV $p\bar p$ collider.

\begin{figure}[htb]
\centerline{
\epsfig{file=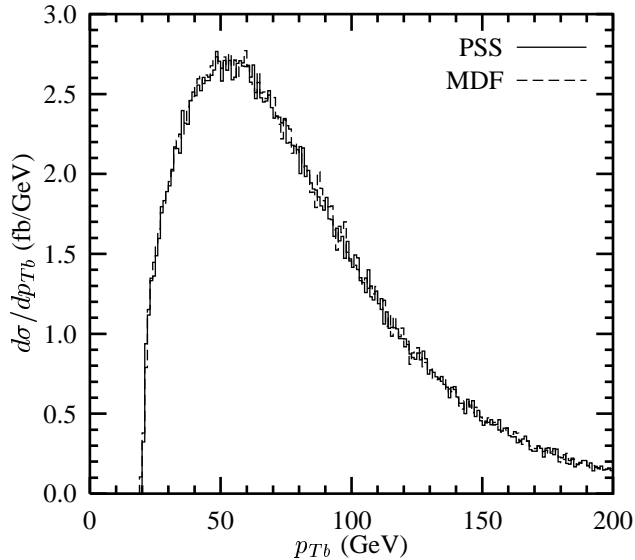, width=3.29in} } 
\caption{\small Transverse momentum distribution of the $b$-jet at NLO
in $s$-channel production of a top quark at the Tevatron ($\sqrt{S}=
2$ TeV) after cuts.  Phase space slicing (PSS) results (solid) and
massive dipole formalism (MDF) results (dashed) are both
shown.\label{fig:ptb}}
\end{figure}

\begin{figure}[htb]
\centerline{
\epsfig{file=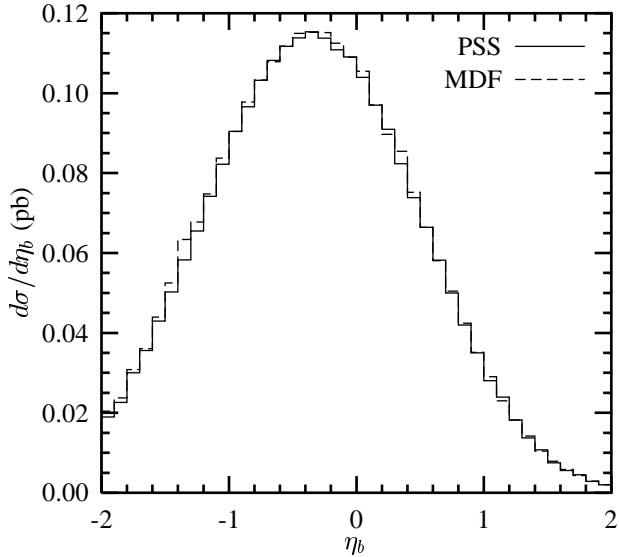, width=3.22in} } 
\caption{\small Pseudorapidity distribution of the $b$-jet at NLO in
$s$-channel production of a top quark at the Tevatron ($\sqrt{S}= 2$
TeV) after cuts.  Phase space slicing (PSS) results (solid) and
massive dipole formalism (MDF) results (dashed) are both
shown.\label{fig:etab}}
\end{figure}

In Figs.~\ref{fig:ptb} and \ref{fig:etab} we present the NLO
transverse momentum $p_{T b}$ and pseudorapidity $\eta_b$
distributions of the $b$-jet in $s$-channel production of a top quark,
with cuts based on the ``jet veto'' search strategy in
Ref.~\cite{Stelzer:1998ni}.  We accept events where only the top-quark
and $b$-jet pass the cuts above, and any additional jets are
either soft ($p_{T j} < 20$ GeV) or are outside the simple detector
($|\eta_j| > 2$).  The phase space slicing method and massive dipole
formalism give identical distributions, even in this region of
restricted phase space.


\begin{figure}[htb]
\centerline{
\epsfig{file=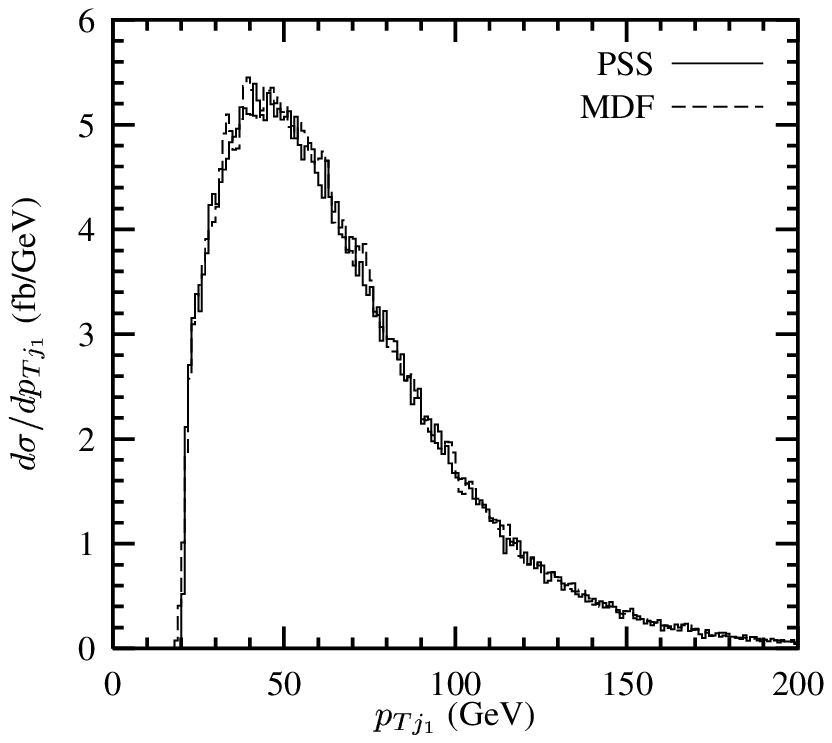, width=3.29in} } 
\caption{\small Transverse momentum distribution of the
highest-$p_T$ jet ($j_1$) at NLO in $t$-channel production of a top
quark at the Tevatron ($\sqrt{S}= 2$ TeV) after cuts.  Phase space
slicing (PSS) results (solid) and massive dipole formalism (MDF)
results (dashed) are both shown.\label{fig:ptj1}}
\end{figure}

\begin{figure}[htb]
\centerline{
\epsfig{file=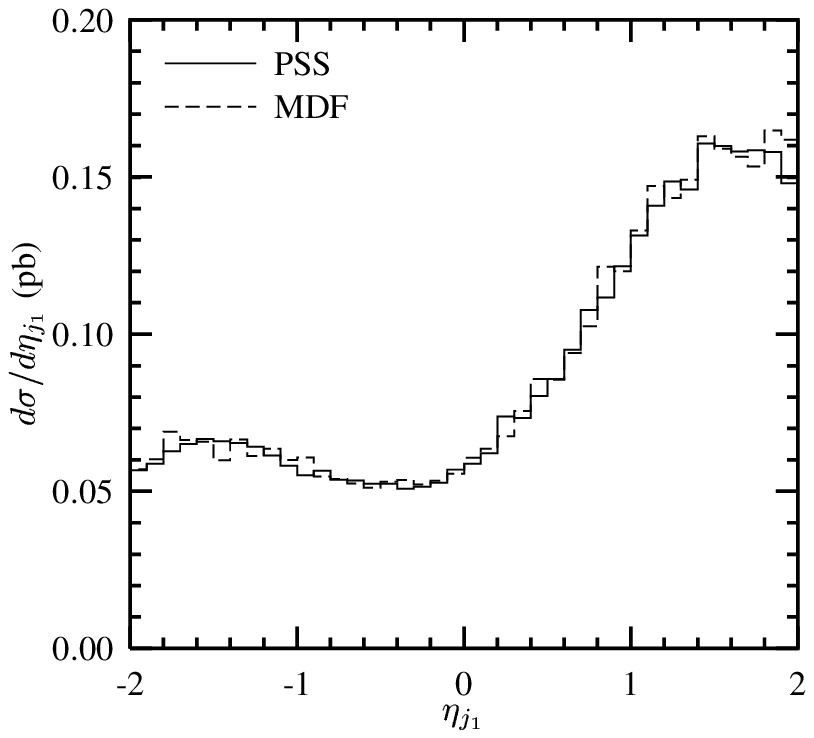, width=3.22in} } 
\caption{\small Pseudorapidity distribution of the highest-$p_T$
jet ($j_1$) at NLO in $t$-channel production of a top quark at
the Tevatron ($\sqrt{S}= 2$ TeV) after cuts.  Phase space slicing
(PSS) results (solid) and massive dipole formalism (MDF) results
(dashed) are both shown.\label{fig:etaj1}}
\end{figure}

In Figs.~\ref{fig:ptj1} and \ref{fig:etaj1} we present the NLO
transverse momentum $p_{T j_1}$ and pseudorapidity $\eta_{j_1}$
distributions of the highest-$p_T$ jet in $t$-channel production of a
top quark, using the same ``jet veto'' search strategy as above.  We
accept events where only the top-quark and highest-$p_T$ jet pass
the cuts above, and any additional jets are either soft ($p_{T j_2} <
20$ GeV) or are outside the simple detector ($|\eta_{j_2}| > 2$).  The
phase space slicing method and massive dipole formalism predict
identical distributions at NLO.


\section{Conclusions}
\label{sec:conclusions}

We present three independent calculations of the fully differential
production of a single top quark plus one jet at next-to-leading order
in hadronic collisions.  At this order in QCD the cross sections
factorize into two non-interfering production modes that may be
identified by the $s$-channel or $t$-channel exchange of a $W$ boson.
The $s$-channel cross section is characterized by having a typical
final state of a top quark and a bottom-quark jet.  The jet in the
$t$-channel cross section tends to be somewhat more forward, and
rarely contains a bottom quark.  New physics scenarios tend to effect
these production modes differently, and hence an accurate measurement
and theoretical calculation of both modes is desirable.  Further, the
CKM matrix element $|V_{tb}|^2$ is directly proportional to the cross
section, and the ability to extract $V_{tb}$ is ultimately limited by
our ability to predict the measured exclusive cross sections with
experimental cuts.

The total cross sections for $s$- and $t$-channel production are
updated with CTEQ5 parton distribution functions for runs I and II of
the Tevatron, and for the LHC.  The $t$-channel cross section is
smaller than previously published by $13\%$.  This decrease is due
entirely to the correction of a bug in the parton distribution
functions.  If the Tevatron continues to run at $1.96$~TeV instead of
2~TeV, the $s$($t$)-channel production modes will produce 4\%(7\%)
fewer events than previously expected.  We discuss some subtleties in
estimating the effects of higher-order corrections on these cross
sections.

We show that the phase space slicing methods of one and two cutoffs
and the massive dipole formalism produce the same jet distributions
and cross sections.  The dipole calculation retains the full spin
dependence of the external particles, and so may be used to predict
spin-dependent correlations of the partons.
We discuss elsewhere \cite{HLPSW2} the effects of
top-quark mass, scales, jet definitions, and parton distribution
functions on the shapes of the NLO distributions.

\subsection*{Acknowledgments}
\label{sec:acknowledgements}

B.H. and Z.S. thank Argonne National Laboratory High Energy Physics 
Division for support during the early stages of the project.  
Their work was supported in part by the U.S. Department of Energy, 
High Energy Physics Division, under contract W-31-109-Eng-38, and
DE-AC02-76CH03000 (at Fermilab).
The work of E.L. and L.P. is supported by the Foundation for 
Fundamental Research of Matter (FOM) and the National Organization 
for Scientific Research (NWO).

\begin{appendix}

\section{Scalar integrals}
\label{sec:scalar_integrals}

We calculate the integrals in $D=4-2\eps$ dimensions and use the notation
\begin{eqnarray} 
\Delta & = & \frac{1}{\eps} - \gamma + \ln 4 \pi \,.
\end{eqnarray}
All integrals are calculated in the Euclidean region (invariants
$p^2<0$ and masses $m^2>0$).  These expressions can then be continued
analytically to the regions of interest, using the substitution $-s
\rightarrow -s-i \eps$ (where $\eps$ denotes a small parameter, not to
be confused with the one appearing in dimensional regularization) and
the formula
\begin{eqnarray}
\ln(-s-i\eps) & = & \ln (|s|) - i \pi \Theta(s) \,.
\end{eqnarray}

Massive integrals: The tadpole is given for $m^2>0$ by
\begin{eqnarray}
A_0(m^2) & =& (4 \pi)^2 \mu^{4-D} \int \frac{d^Dk}{(2\pi)^D i}
\frac{1}{k^2-m^2} 
= - m^2 \Gamma(-1+\eps) (4\pi)^\eps \left( \frac{m^2}{\mu^2} \right)^{-\eps}
\nonumber \\
& = & m^2 \left( \Delta + 1 - \ln \frac{m^2}{\mu^2} \right) + O(\eps) \,.
\end{eqnarray}
The bubble with one internal mass is given for $-p^2>0, m^2>0$ by
\begin{eqnarray}
B_0^{(a)}(p^2,m^2) & =& (4 \pi)^2 \mu^{4-D} \int \frac{d^Dk}{(2\pi)^D i}
\frac{1}{k^2 ((k-p)^2-m^2)} \nonumber \\
& = & \Delta + 2 - \ln \frac{m^2-p^2}{\mu^2} + \frac{m^2}{-p^2}
\ln \frac{m^2}{m^2-p^2} + O(\eps) \,.
\end{eqnarray}

On-shell integrals:
The bubble with one internal mass is given for $p^2=m^2>0$ by
\begin{eqnarray}
B_0^{(b)}(m^2) & =& (4 \pi)^2 \mu^{4-D} \int \frac{d^Dk}{(2\pi)^D i}
\frac{1}{k^2 ((k-p)^2-m^2)} \nonumber \\
& = & \Delta + 2 - \ln \frac{m^2}{\mu^2} + O(\eps) \,.
\end{eqnarray}
The triangle with two external masses and one internal mass, between
the two external massive lines is given for $p_1^2=0$,
$(p_1+p_2)^2=m^2>0$ by
\begin{eqnarray}
\lefteqn{
C_0^{(a)}(p_2^2,m^2) = (4 \pi)^2 \mu^{4-D} \int \frac{d^Dk}{(2\pi)^D i} 
\frac{1}{k^2 (k-p_1)^2 ((k-p_1-p_2)^2-m^2)} } & & \nonumber \\
& = & \left( 4 \pi \mu^2 \right)^\eps \frac{\Gamma(1+\eps)}{m^2-p_2^2}
 \left\{ -\frac{1}{2\eps^2} + \frac{1}{\eps} \left( \ln\left(m^2-p_2^2\right)
- \frac{1}{2} \ln(m^2) \right)  \right. \nonumber \\
& & \left.
 + \frac{1}{4} \ln^2(m^2) - \frac{1}{2} \ln^2\left(m^2-p_2^2\right) 
 +\;\mbox{Li}_2\left(\frac{-p_2^2}{m^2-p_2^2}\right) \right\} +O(\eps)
\nonumber \\
& = & 
 \frac{1}{m^2-p_2^2} \left\{
  -\frac{1}{2} \Delta^2 + \Delta \left( -\frac{1}{2} \gamma + \frac{1}{2}
\ln 4 \pi 
                                        + \frac{1}{2} \ln \frac{m^2}{\mu^2}
                                        + \ln \frac{m^2-p_2^2}{m^2} \right)
 - \frac{1}{4} \left( \gamma -\ln 4 \pi \right)^2 
               \right. \nonumber \\
& & \left.
 + \frac{1}{4} \ln^2 \left( \frac{m^2}{\mu^2} \right)
 - \frac{1}{2} \ln^2 \left( \frac{m^2-p_2^2}{\mu^2} \right)
 + \; \mbox{Li}_2\left(\frac{-p_2^2}{m^2-p_2^2}\right) 
 - \frac{\pi^2}{24} \right\} + O(\eps) \,.
\end{eqnarray}

Integrals with no internal masses: The bubble with $-p^2>0$ is given by
\begin{eqnarray}
B_0^{(c)}(p^2) & =& (4 \pi)^2 \mu^{4-D} \int \frac{d^Dk}{(2\pi)^D i}
\frac{1}{k^2 (k-p)^2} 
 = \Gamma(\eps) \frac{\Gamma^2(1-\eps)}{\Gamma(2-2\eps)} (4\pi)^\eps \left(
\frac{-p^2}{\mu^2} \right)^{-\eps} \nonumber \\
& = & \Delta + 2 - \ln \frac{-p^2}{\mu^2} + O(\eps) \,.
\end{eqnarray}
The triangle with one external mass $p_1^2=p_2^2=0$, $-s=-2p_1p_2>0$
is given by
\begin{eqnarray}
\lefteqn{
C_0^{(b)}(s) = (4 \pi)^2 \mu^{4-D} \int \frac{d^Dk}{(2\pi)^D i} 
\frac{1}{k^2 (k-p_1)^2 (k-p_1-p_2)^2} } & & \nonumber \\
& = & \frac{1}{s} \frac{1}{\eps^2} \frac{\Gamma(1+\eps) \Gamma^2(1-
\eps)}{\Gamma(1-2\eps)}
(4 \pi)^\eps \left( \frac{-s}{\mu^2} \right)^{-\eps} \nonumber \\
& = & 
\frac{1}{s} \left( \Delta^2 + \Delta \left( \gamma - \ln 4 \pi - \ln
\frac{-s}{\mu^2} \right)
+ \frac{1}{2} \left( \gamma - \ln 4 \pi \right)^2 
+ \frac{1}{2} \ln^2\left( \frac{-s}{\mu^2} \right)
- \frac{\pi^2}{12} \right) + O(\eps) \,. \nonumber \\
\end{eqnarray}

Six-dimensional integrals: The triangle with an additional power of
$k_{-2\eps}^2$ in the numerator is equivalent to an integral in
$D=6-2\eps$ dimensions and yields for the massless and massive case
\begin{eqnarray}
C_0^{(-2\eps)} & =& (4 \pi)^2 \mu^{4-D} \int \frac{d^Dk}{(2\pi)^D i} 
\frac{k_{(-2\eps)}^2}{k^2 (k-p_1)^2 ((k-p_1-p_2)^2-m^2)} \nonumber \\
& = & - \frac{1}{2} + O(\eps) \,.
\end{eqnarray}

\end{appendix}

\end{document}